\documentclass{aa}
\usepackage{xeCJK}
\usepackage{graphicx}
\usepackage[colorlinks=true, allcolors=blue]{hyperref}
\usepackage{amsmath}
\usepackage{amssymb}
\usepackage[nameinlink,capitalize]{cleveref}
\usepackage{mathrsfs}
\usepackage{natbib}
\usepackage{booktabs}
\usepackage{longtable}
\usepackage{array}
\usepackage{orcidlink}
\usepackage{xspace}

\newcommand{\trinity}{\textsc{Trinity}\xspace}
\newcommand{\ccm}{\mathrm{cm^{-3}}}
\newcommand{\pc}{\mathrm{pc}}
\newcommand{\kms}{\mathrm{km\,s^{-1}}}
\newcommand{\Myr}{\mathrm{Myr}}
\newcommand{\hii}{H~\textsc{ii}\xspace}
\newcommand{\Msun}{\mathrm{M}_\odot}
\newcommand{\kB}{k_\mathrm{B}}
\newcommand{\mH}{m_\mathrm{H}}
\newcommand{\xHe}{x_{\mathrm{He}}}
\newcommand{\mup}{\mu_\mathrm{p}}
\newcommand{\mun}{\mu_\mathrm{n}}
\newcommand{\muH}{\mu_\mathrm{H}}
\newcommand{\ZHe}{Z_\mathrm{He}}

\newcommand{\Rts}{R_\mathrm{ts}}           % wind Termination Shock  (was R_1)
\newcommand{\Rb}{R_\mathrm{b}}             % Bubble outer boundary   (was R_2)
\newcommand{\Rif}{R_\mathrm{if}}           % Ionisation Front
\newcommand{\Rshell}{R_\mathrm{sh}}        % Shell outer edge         
\newcommand{\Rcloud}{R_\mathrm{cloud}}     % Cloud boundary          (was R_cl)
\newcommand{\Rcore}{R_\mathrm{core}}       % Core radius             (was R_c / r_c)
\newcommand{\vb}{v_\mathrm{b}}           % Bubble velocity          (was v_2)
\newcommand{\vw}{v_\mathrm{w}}             % Wind terminal speed
\newcommand{\vsn}{v_\mathrm{SN}}           % SN ejecta speed
\newcommand{\Mcloud}{M_\mathrm{cloud}}     % Cloud mass              (was M_cl)
\newcommand{\Mgas}{M_\mathrm{gas}}
\newcommand{\Mstar}{M_\star}               % Cluster mass
\newcommand{\Msh}{M_\mathrm{sh}}           % Swept-up shell mass

\newcommand{\Mdotw}{\dot{M}_\mathrm{w}}   % Wind mass-loss rate
\newcommand{\Mdotsn}{\dot{M}_\mathrm{SN}} % SN ejecta mass rate
\newcommand{\pdotw}{\dot{p}_\mathrm{w}}   % Wind momentum injection rate

\newcommand{\Eb}{E_\mathrm{b}}             % Bubble thermal energy
\newcommand{\Pb}{P_\mathrm{b}}             % Bubble thermal pressure (was also P_b^th)
\newcommand{\Vb}{V_\mathrm{b}}             % Bubble volume
\newcommand{\Tb}{T_\mathrm{b}}             % Bubble temperature at R_b
\newcommand{\fesc}{f^{\mathrm{LyC}}_\mathrm{esc}}
\newcommand{\fabs}{f_\mathrm{abs}}
\newcommand{\Pdrive}{P_\mathrm{drive}}     % Net driving pressure
\newcommand{\PHII}{P_{\mathrm{H\,\textsc{ii}}}}  % Photoionised gas pressure
\newcommand{\Pram}{P_\mathrm{ram}}         % Ram pressure
\newcommand{\Pext}{P_\mathrm{ext}}         % External confining pressure
\newcommand{\PISM}{P_\mathrm{ISM}}         % Ambient ISM floor       (was P_0)

\newcommand{\Frad}{F_\mathrm{rad}}         % Radiation force
\newcommand{\Fgrav}{F_\mathrm{grav}}       % Gravity                 (was F_G)
\newcommand{\Fram}{F_\mathrm{ram}}         % Ram momentum flux
\newcommand{\Tion}{T_\mathrm{ion}}         % Ionised gas temp ~10^4 K  (was T_i)
\newcommand{\Tneu}{T_\mathrm{neu}}         % Neutral shell temp        (was T_a)

\newcommand{\rhocore}{\rho_\mathrm{core}}  % Core density              (was \rho_c)
\newcommand{\ncore}{n_\mathrm{core}}       % Core number density
\newcommand{\nshb}{n_{\mathrm{sh}}(\Rb)}       % Shell density at R_b (inner edge)   (was n_b / n_0)
\newcommand{\nshifStr}{n_\mathrm{sh,if}^{\,\mathrm{Str}}}  % n_{sh,if} from Stroemgren bal.
\newcommand{\alpharho}{\alpha_\rho}        % Density slope  (was bare α in earlier drafts)

\newcommand{\Lw}{L_\mathrm{w}}             % Wind mechanical luminosity
\newcommand{\Lmech}{L_\mathrm{mech}}       % Total mechanical luminosity
\newcommand{\Lcool}{L_\mathrm{cool}}       % Radiative cooling luminosity
\newcommand{\Lnet}{\Lambda_\mathrm{net}}   % Net volumetric cooling function
\newcommand{\Qi}{Q_\mathrm{i}}             % Ionising photon rate

\newcommand{\tfe}{t_\mathrm{fe}}     % free expansion time             (was t_0)
\newcommand{\tcool}{t_\mathrm{cool}}       % Cooling timescale
\newcommand{\tsc}{t_\mathrm{sc}}           % Sound-crossing time
\newcolumntype{L}[1]{>{\raggedright\arraybackslash}p{#1}}

\begin{document}

\title{TRINITY: A coupled model of winds, radiation, and photoionised gas in molecular clouds}
\subtitle{I. Methods and validation}

\author{
  Jia~Wei~Teh (郑家伟)\inst{1}\thanks{\email{jiaweiteh.astro@gmail.com}}\orcidlink{0000-0001-7863-5047}
  \and Ralf~S.~Klessen\inst{1,2}\orcidlink{0000-0002-0560-3172}
  \and Simon~C.~O.~Glover\inst{1}\orcidlink{0000-0001-6708-1317}
  \and Kathryn~Kreckel\inst{3}\orcidlink{0000-0001-6551-3091}
}

\institute{
  Universit\"at Heidelberg, Zentrum f\"ur Astronomie, Institut f\"ur Theoretische Astrophysik, Albert-Ueberle-Stra\ss e 2, 69120 Heidelberg, Germany
  \and
  Universit\"at Heidelberg, Interdisziplin\"ares Zentrum f\"ur Wissenschaftliches Rechnen, Im Neuenheimer Feld 205, 69120 Heidelberg, Germany
  \and
  Universit\"at Heidelberg, Zentrum f\"ur Astronomie, Astronomisches Rechen-Institut, M\"onchhofstra\ss e 12-14, 69120 Heidelberg, Germany
}

\abstract
% context
{Multi-wavelength surveys place cloud dispersal at $\sim1$--$5\,\Myr$ after massive-star emergence, before the first core-collapse supernovae. Whether a cloud disperses, re-collapses, or leaks Lyman-continuum (LyC) photons depends on how pre-supernova winds, radiation pressure, and photoionised gas pressure ($\PHII$) couple to the swept-up shell.}
% aims
{We introduce \textsc{Trinity}, a 1D thin-shell code that succeeds \textsc{warpfield} and follows stellar feedback across varied initial cloud structures.}
% methods
{\textsc{Trinity} evolves the bubble-shell structure under winds, supernovae, direct and dust-reprocessed radiation pressure, $\PHII$, and gravity. A phase-aware driving prescription uses higher values of the hot-bubble and photoionised gas pressures in the energy-driven phase and accounts for $\PHII$ and direct ram pressure in the momentum-driven phase. The initial cloud may be uniform, a piecewise power law, or a Bonnor--Ebert sphere. Shell structure, hot-bubble cooling, photon absorption, and LyC escape are solved alongside the dynamics.}
% results
{We validated the code against analytic wind and photoionisation limits and explored runs with cloud masses $\Mcloud=10^5$--$10^{6.5}\,\Msun$, core densities $\ncore=10^3$--$10^4\,\ccm$, and star-formation efficiencies $\varepsilon=0.01$--0.30.
Including $\PHII$ increases the bubble radius by $\sim10\%$ at $10\,\Myr$ in the fiducial validation run.
In a higher-efficiency example, the energy-driven phase lasts on the order of $1\,\Myr$, radiation pressure remains subdominant, and $\PHII$ remains dynamically significant after the transition to momentum-dominated evolution.
Density structure changes both phase durations and outcomes. At fixed $\Mcloud$, $\ncore$, and $\varepsilon$, homogeneous and shallow profiles re-collapse, while a steep $\rho\propto r^{-2}$ cloud keeps expanding.
In a dispersing grid, Bonnor--Ebert clouds reach their dispersal time ($\tau_{\mathrm{disp}}$) much later than homogeneous clouds.}
% conclusions
{The first results show that both $\PHII$ and cloud structure are important for a complete dynamical treatment: $\PHII$ alters late-time shell expansion, and density structure shifts dispersal thresholds and emergence times for a fixed stellar population.
\textsc{Trinity} provides an efficient framework for mapping feedback dominance across a cloud parameter space and produces the diagnostics needed to compare with resolved \hii region observations.}

\keywords{methods: numerical -- ISM: bubbles -- ISM: clouds -- ISM: H~\textsc{ii} regions -- ISM: kinematics and dynamics -- stars: formation}

\authorrunning{J.\,W.\,Teh et al.}
\titlerunning{\trinity I: Unified feedback in bubbles}
\maketitle

\section{Introduction}
\label{sec:intro}

Giant molecular clouds (GMCs) form stars inefficiently. 
Across nearby galaxies, integrated cloud-scale star-formation efficiencies range from roughly a few to ten per cent, and parent clouds disperse within $1$--$5\,\Myr$ once massive stars emerge \citep{2020MNRAS.493.2872C,2022MNRAS.516.3006K}. 
The first core-collapse supernovae (SNe) arrive only after $\sim 3$--$4\,\Myr$, hinting that cloud clearing is caused by pre-SN feedback (e.g. photoionisation, stellar winds, and radiation pressure; see \citealt{2014PhR...539...49K,2015NewAR..68....1D, 2018A&A...611A..70R}). 
The remaining question is which of these channels dominates, when, and for which cloud properties.

Multi-wavelength surveys time cloud clearing by matching molecular gas, young stars, dust emission, and ionised gas at cloud-scale resolution. 
The Physics at High Angular resolution in Nearby GalaxieS (PHANGS) survey provides the clearest example: ALMA traces the cold molecular reservoir; HST resolves the exposed young clusters; JWST recovers embedded sources and dust structure; and VLT/MUSE measures the nebular gas \citep{2021ApJS..257...43L,2022ApJS..258...10L,2022A&A...659A.191E,2023ApJ...944L..17L}.
Resolved CO, H$\alpha$, and $24\,\mu$m analyses place the duration of the embedded phase of massive star formation at $2$--$7\,\Myr$ \citep{2021MNRAS.504..487K} and demonstrate that GMCs typically disperse within $\sim3\,\Myr$ once massive stars appear \citep{2022MNRAS.509..272C}.
PHANGS-JWST extends the same argument into the dust-obscured phase. Statistical timing across 37 galaxies puts the embedded-to-exposed transition below $4\,\Myr$ and below $1\,\Myr$ in some environments \citep{2026A&A...706A.186R}.
The H$\alpha$ morphologies provided by the Legacy ExtraGalactic UV Survey (LEGUS; \citealt{2015AJ....149...51C}) give consistent clearing times of roughly $1$--$2\,\Myr$ for the most massive clusters \citep{2022MNRAS.512.1294H}. 
In the Milky Way, the SDSS-V Local Volume Mapper now brings comparable optical emission-line diagnostics to parsec and sub-parsec scales over large sky areas \citep{2024A&A...689A.352K,2024AJ....168..198D, 2026A&A...706A..81S}. 
These measurements indicate the same ordering: most regions expose their massive stars before the first core-collapse SN arrives.

Cloud dispersal time is not universal. More massive clusters emerge from their natal gas earlier than lower-mass ones \citep{Pedrini2026}. 
\citet{2024ApJ...967..102M} finds that low-mass clusters in NGC~4449 remain embedded for $5$--$6\,\Myr$, while more massive clusters expose their surroundings by $\sim4\,\Myr$. 
The mechanism behind cloud clearing also changes with scale. In the massive, high-surface-density cloud models \citep{Polak2024}, collapse converts most of the gas into stars before feedback can halt star formation.
\citet{2021MNRAS.508.5362B} analysed $\sim5800$ \hii regions across 19 galaxies and find that most are overpressured, depending on the assumed density structure. 
Direct radiation pressure primarily dominates in the most compact, embedded, or luminous environments \citep[see e.g.][]{2014ApJ...795..121L}. For instance, in 30~Doradus it exceeds photoionised gas pressure within $\sim75\,\mathrm{pc}$ of the central cluster \citep{2011ApJ...731...91L}. By contrast, in Galactic regions with $r\lesssim0.5\,\mathrm{pc}$, dust-reprocessed infrared radiation pressure dominates roughly $84\%$ of cases \citep{2021ApJ...908...68O}.
Beyond the compact phase, photoionised gas pressure usually dominates \citep{2025ApJ...982..140P}. 
Resolved studies that connect pressure terms to candidate ionising stellar populations recover this hierarchy and begin to map its dependence on metallicity, cluster mass, and the evolutionary state \citep{2022A&A...662L...6B,2023MNRAS.522.2369S,2024A&A...685A..46R,2026A&A...706A..95B,2026A&A...708A.202S}. 
Direct cross-matching of star-cluster catalogues with H~\textsc{ii}-region catalogues further constrains how many ionising photons leak into the diffuse ionised gas \citep{2023MNRAS.524.1191T,2026A&A...708A.202S}.

The target is therefore no longer the cloud lifetime alone, but how much feedback couples to the natal cloud, how much leaks out, and what medium the first SN encounters.
The same pre-SN clearing physics sets the environment into which SNe explode \citep{2023ApJ...944L..24W, 2023ApJ...944L..22B}. 
Radiation-hydrodynamic calculations show that winds and photoionisation carve low-density cavities tens of parsecs across, reducing the density of the surrounding interstellar medium (ISM; \citealt{2024A&A...690A..72F}). 
If the expanding cavity has already reached the edge of the cloud, the SN expands in a semi-confined fashion through low-density channels rather than spherically into an intact cloud. This can amplify small-scale turbulence \citep{2025MNRAS.540.1124L}. 
Modelling the pre-SN phase correctly is therefore a precondition for modelling SN feedback at all.

Existing tools cover different slices of this problem. Semi-analytic codes such as \textsc{warpfield} \citep{2017MNRAS.470.4453R,2019MNRAS.483.2547R} follow winds, radiation pressure, SNe, cooling, and gravity on cloud scales and can survey broad parameter ranges efficiently.
\textsc{warpfield-emp} couples \textsc{warpfield} to \textsc{cloudy} \citep{2017RMxAA..53..385F} and \textsc{polaris} \citep{2016A&A...593A..87R} to predict emission lines from evolving \hii regions \citep{2020MNRAS.496..339P}, while \textsc{toddlers} \citep{2023MNRAS.526.3871K,2024A&A...692A..79K} computes emission libraries for star-forming regions by coupling similar evolution models to the radiative transfer code \textsc{skirt} \citep{2011ApJS..196...22B}. 
\citet{2026A&A...709A.100J} include the metallicity dependence in the wind--cooling balance, identifying a standing-shell regime in low-metallicity clouds. Three-dimensional simulations follow feedback with similar physical fidelity but at a much higher computational cost \citep[see][]{2015MNRAS.454..238W,2016MNRAS.456.3432G,2020MNRAS.492.1594S, 2021MNRAS.506.2199G}.

The parameter space of the problem is large. The cloud mass, mean density, density profile, and cluster mass can all have a significant impact on the evolution of the feedback-driven cavity. 
Mapping this parameter space statistically requires an efficient forward model that self-consistently follows every form of pre-SN feedback and produces outputs that directly align with resolved observations.

Thus, \trinity is designed to couple winds, SNe, direct and dust-reprocessed radiation pressure, photoionised gas pressure ($\PHII$), and gravity through a continuous energy- to momentum-driven bubble evolution. 
In this first paper of the \trinity series, we introduce two changes relative to \textsc{warpfield}.
First, $\PHII$ enters the shell equation of motion explicitly and combines with the hot-bubble pressure, $\Pb$. 
Second, the initial cloud is not restricted to simple homogeneous density profiles. Uniform, power-law, and Bonnor--Ebert (BE) structures can be compared at different cloud masses and star-formation efficiencies. 

This study is structured as follows: Sect.~\ref{sec:methods} describes the \trinity\ model. 
Section~\ref{sec:validation} validates against analytic limits; Sect.~\ref{sec:results} demonstrates the diagnostic suite; and Sect.~\ref{sec:caveats} collects the limitations before Sect.~\ref{sec:conclusions} concludes.

\begin{figure*}
    \centering
    \includegraphics[width=\textwidth]{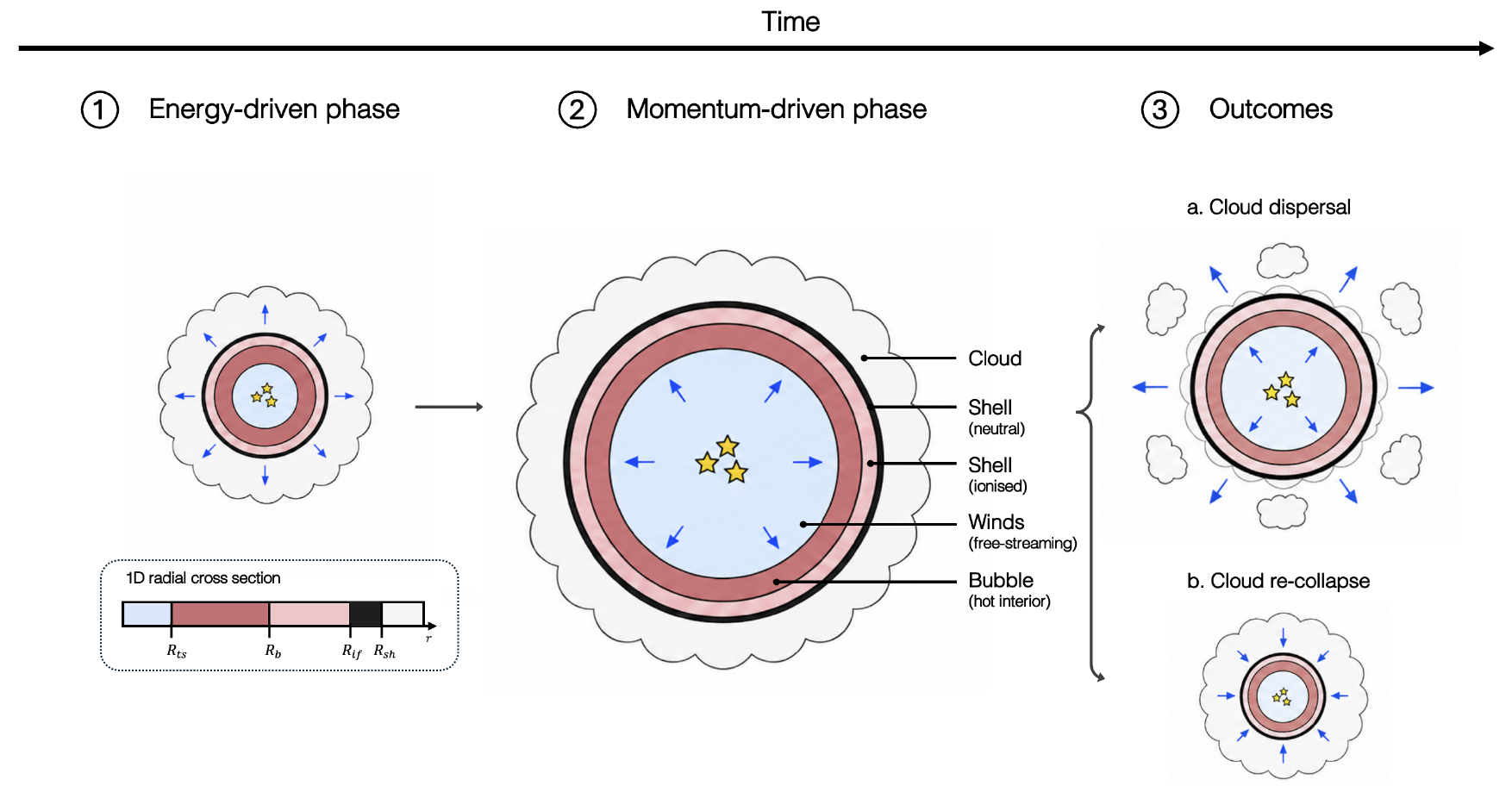}
    \caption{Schematic of the bubble geometry and evolutionary sequence modelled by \trinity. The central cluster (yellow stars) injects free-streaming winds (light blue), which thermalise at the wind termination shock, $\Rts$. 
    The shocked wind fills the hot bubble out to the bubble radius, $\Rb$ (red). Outside $\Rb$, the swept-up shell contains an ionised layer (pink) bounded by the ionisation-front radius, $\Rif$, and a neutral layer (black) extending to the outer shell radius, $\Rshell$. The surrounding grey region denotes the natal cloud. 
    Inset: Corresponding 1D radial structure, ordered as $\Rts < \Rb < \Rif < \Rshell$. 
    In the energy-driven phase (left), the hot-bubble pressure dominates the shell dynamics and the classical wind-bubble solution gives $R\propto t^{3/5}$ in a uniform medium \citep{1977ApJ...218..377W}. 
    As cooling drains the hot bubble, the system enters the momentum-driven phase (centre), where shell expansion is supported by the photoionised gas pressure, radiation pressure, and direct ram pressure from winds and SN ejecta. 
    The final outcome depends on whether feedback overcomes gravity. The shell may break through and disperse the cloud (top right) or decelerate, turn around, and re-collapse (bottom right). The blue arrows indicate the direction of shell motion.}
    \label{fig:schematic}
\end{figure*}

\section{The TRINITY model}
\label{sec:methods}

\textsc{Trinity}\footnote{\url{https://jiaweiteh.github.io/trinity-web/}} is a 1D, spherically symmetric thin-shell code\footnote{We treated the shell as a single mass element in the equation of motion, with a radius and velocity.} that follows the expansion of a feedback-driven bubble and swept-up shell into a GMC. It builds on the \textsc{warpfield} framework \citep{2017MNRAS.470.4453R, 2019MNRAS.483.2547R}, with the extensions noted in Sect.~\ref{sec:intro}: an explicit photoionised gas pressure term in the shell equation of motion, a phase-aware combination of $\PHII$ and the bubble pressure across the energy- to momentum-driven transition, and support for BE profiles alongside uniform and power-law clouds. The code solves a coupled system of ordinary differential equations for the bubble radius $\Rb$, bubble velocity $\vb$, and bubble thermal energy $\Eb$, subject to forces from stellar winds, SNe, direct and dust-reprocessed radiation pressure, photoionised gas pressure, and self-gravity of the shell and enclosed cluster. At each time step, a hydrostatic shell-structure calculation determines the ionisation-front position, absorbed radiation fraction, and infrared optical depth, so that the coupling between radiation and gas is solved self-consistently rather than prescribed.

\trinity is written in \textsc{Python} and uses the \textsc{SciPy} \texttt{solve\_ivp} interface to the \textsc{LSODA} stiff integrator \citep{Hindmarsh1983, Petzold1983}. Stellar wind luminosities, mass-loss rates, ionising photon rates, and bolometric luminosities are drawn from \textsc{starburst99} population synthesis tables \citep{1999ApJS..123....3L}, though the modular input interface accepts any tabulated stellar-evolution model (see Appendix~\ref{app:stellar_feedback_table}). 
A single run for a $10^5\,\Msun$ cloud with a $10^3\,\Msun$ cluster completes in $\sim 30$ minutes on one core, fast enough for systematic surveys in cloud mass, density profile, star formation efficiency (SFE), and ISM conditions. The code is released under an open-source licence; installation instructions and example scripts are provided in the repository.

\subsection{Initial conditions}
\label{sec:initial-conditions}

A star cluster of mass $\Mstar = \varepsilon\,\Mcloud$ forms instantaneously\footnote{We placed the cluster on the zero-age main sequence (ZAMS). See Appendix~\ref{app:ZAMS} for the effect of neglecting pre-main-sequence evolution.} at the centre of a spherical molecular cloud of initial mass ($\Mcloud$), SFE ($\varepsilon$), and metallicity ($Z$). 
The remaining cloud mass is $\Mgas = (1-\varepsilon)\Mcloud$, with radial density profile ($\rho_\mathrm{cloud}(r)$; see Sect.~\ref{sec:dens_main}). 
The cluster immediately launches a stellar wind, which must push through this gas before any bubble-shell structure can form.

We adopted a helium-to-hydrogen number ratio $\xHe = 0.1$ throughout, giving a mass per hydrogen nucleus $\muH=1.4\,\mH$, where $\mH$ is the proton mass. 
The mean mass per particle depends on the ionisation state: $\mu = \mun$ in neutral gas and $\mu = \mup$ in ionised gas, with
\begin{align}
    \mun &= \frac{\muH}{1 + \xHe} = \frac{14}{11}\,\mH, \label{eq:mun} \\
    \mup &= \frac{\muH}{2 + \xHe(1 + \ZHe)} = \frac{14}{23}\,\mH, \label{eq:mup}
\end{align}
where $\ZHe = 2$ throughout this work (helium doubly ionised)\footnote{We treated helium as doubly ionised throughout the hot wind-blown bubble and swept-up shell for consistency with previous semi-analytic feedback codes \citep{2014ApJ...785..164M,2017MNRAS.470.4453R}. The doubly ionised assumption is precise for the bubble but an approximation for the $T\sim10^4\,\mathrm{K}$ shell, where helium is mostly singly ionised.}.
All following number densities, $n$, denote hydrogen nuclei densities.

The cloud boundary $\Rcloud$ is defined implicitly by
\begin{equation}
  \Mgas = 4\pi\int_0^{\Rcloud} r^2\,\rho_\mathrm{cloud}(r)\,\mathrm{d}r.
  \label{eq:cloud_mass}
\end{equation}
Thus, $\Rcloud$ is a derived consequence of the chosen density profile, and $\Mgas$ is not a free parameter.

The cluster injects mass at rate $\Mdotw$ with terminal speed $\vw$, so the mechanical luminosity and momentum injection rate are
\begin{equation}
  \Lw = \tfrac{1}{2}\Mdotw\vw^2, \qquad \pdotw = \Mdotw\vw.
  \label{eq:Lw_pdot}
\end{equation}
In practice, $\Lw(t)$ and $\pdotw(t)$ are time-dependent quantities read from stellar evolution tables (see Appendix~\ref{app:stellar_feedback_table}). The expressions above define instantaneous values at any given $t$.

Initially, the wind expands freely into the cloud. The shell becomes dynamically important once the swept-up mass equals the ejected wind mass, which defines the free-expansion time
\begin{equation}
  \tfe = \sqrt{\frac{3}{4}\frac{\Mdotw}{\pi\rhocore\vw^3}},
  \label{eq:tfe}
\end{equation}
where $\rhocore$ is the core density of the cloud \citep[see][]{1999isw..book.....L}. 
For our parameters $\tfe \sim 10^3$--$10^4$\,yr, far shorter than any feedback-relevant timescale, so the initial conditions set at $\tfe$ are insensitive to the detailed free-expansion dynamics.

At $t = \tfe$, the bubble structure is already self-similar, so the initial state of the bubble-shell system follows directly from a dimensional analysis of the energy-driven solution:
\begin{align}
  \vb &= \frac{2\Lw}{\pdotw}, \label{eq:vsh_init} \\[4pt]
  \Rb  &= \vb\,\tfe, \label{eq:Rb_init} \\[4pt]
  \Eb  &= \frac{5}{11}\,\Lw\,\tfe. \label{eq:Eb_init}
\end{align}
Equation~\eqref{eq:Eb_init} is the adiabatic energy content of a Weaver bubble at $\tfe$, where the $5/11$ factor arises from the $\gamma = 5/3$ self-similar profile \citep{1977ApJ...218..377W}.
The interior temperature profile at $t = \tfe$ follows
\begin{equation}
  T(\zeta) = 1.51\times10^6\, L_{w,36}^{8/35}\,\ncore^{2/35}\,t_6^{-6/35}\,(1-\zeta)^{2/5} \;\mathrm{K},
  \label{eq:Tbubble}
\end{equation}
where $\zeta \equiv r/\Rb \in [0,1]$, $L_{w,36} \equiv \Lw / 10^{36}\,\mathrm{erg\,s^{-1}}$, and $t_6 \equiv \tfe / 10^6\,\mathrm{yr}$. We evaluate $T$ at $\zeta = 0.9$, near the outer bubble boundary, to characterise the conduction front that drives mass evaporation into the hot interior; this value initialises the temperature solver for the energy-driven phase. 
Together, Eqs. \eqref{eq:cloud_mass}--\eqref{eq:Tbubble} fully specify the state vector $(\Rb, \vb, \Eb, \Tb)$ at $t = \tfe$ and define the cloud geometry into which the bubble expands.

 \subsection{Bubble structure}
\label{sec:bubble-structure}

A wind-blown bubble has four zones (Fig.~\ref{fig:schematic}; \citealt{1975ApJ...200L.107C, 1977ApJ...218..377W}): (a)~a freely streaming hypersonic stellar wind out to the wind termination shock at $\Rts$;
(b)~the hot bubble --- a region of shocked stellar wind ($T \sim 10^7$\,K) bounded by $\Rts$ and the outer contact discontinuity at $\Rb$;
(c)~a thin, dense, cold shell of swept-up ISM; and 
(d)~the undisturbed GMC or ISM. The hot bubble is collisionally ionised at $T\sim10^6$--$10^8$\,K, so it contains essentially no neutral hydrogen and is transparent to Lyman-continuum (LyC) photons.
Ionising radiation therefore reaches the swept-up shell at $\Rb$ unattenuated, and the photoionised region always surrounds the bubble, extending from $\Rb$ outwards to the ionisation front at $\Rif\geq \Rb$.

The bubble evolves through three dynamical phases depending on whether its internal energy is conserved or radiated away \citep{1977ApJ...218..377W, 1988ApJ...324..776M}: an early energy-driven phase in which the overpressured interior drives the shell ($R \propto t^{3/5}$);
a transition phase when radiative losses drain the thermal reservoir on a sound-crossing timescale;
and a momentum-driven phase in which the shell coasts on the accumulated momentum of winds and SNe (see Sect.~\ref{sec:transition}).
Density gradients and photoionised gas pressure shift the timing and sharpness of each transition in real clouds \citep{2017MNRAS.470.4453R, 2019MNRAS.483.2547R}. 

With the initial state $(\Rb, \vb, \Eb, \Tb)$ established at $\tfe$, we advance this four-zone structure through two coupled equations: one for the bubble's momentum,
\begin{align}
    \frac{\mathrm{d}}{\mathrm{d}t}\!\left(\Msh\dot{\Rb}\right) &= 4\pi \Rb^2\!\left(\Pdrive - \Pext\right) + \Frad - \Fgrav\;, \label{eq:motion} 
\end{align}
and one for the bubble's thermal energy,
\begin{align}
    \frac{\mathrm{d}\Eb}{\mathrm{d}t} &= \Lmech - \Lcool - 4\pi \Rb^2\,\Pb\,\dot{\Rb}\;, \label{eq:energy}
\end{align}
where $\Msh$ is the swept-up shell mass and $\Lmech$ is the total mechanical luminosity injected by winds and SNe.

Let $\Vb \equiv 4\pi/3(\Rb^3 - \Rts^3)$ denote the volume of the hot bubble between the wind termination shock at $\Rts$ and the contact discontinuity at $\Rb$. The bubble thermal pressure follows
\begin{equation}
    \Pb = \frac{(\gamma - 1)\,\Eb}{\Vb},
    \label{eq:Pb}
\end{equation}
with $\gamma = 5/3$ for an ideal monatomic gas. 
The wind termination shock radius $\Rts$, where the free-streaming wind thermalises, is set by the pressure balance between the ram pressure at $\Rts$ and the hot bubble:
\begin{equation}
    \Rts = \left[\frac{3\Fram\,\Vb}{8\pi\,\Eb}\right]^{1/2},
    \label{eq:Rts}
\end{equation}
where $\Fram \equiv \Mdotw\vw + \Mdotsn\vsn$ is the total mechanical momentum injection rate from stellar winds and SNe combined. 
Since $\Vb$ itself depends on $\Rts$, Eq.~\eqref{eq:Rts} is implicit and is solved numerically at each time step given the current state $(\Rb, \Eb)$.

The mass flux from the shell into the hot bubble is set by conductive evaporation at the inner shell edge \citep{1977ApJ...211..135C, 1977ApJ...218..377W}.  The evaporation rate is determined by the bulk inflow velocity of shell material heated by the conduction front, evaluated at $r \to R_\mathrm{b}$:
\begin{equation}
    \dot{M}_\mathrm{b} = \lim_{r \to \Rb} 4\pi r^2 \frac{\mu\,\Pb}{\kB} \frac{\alpha r/t - \vb}{T},
    \label{eq:Mbdot}
\end{equation}
where $\alpha \equiv \mathrm{d}\ln\Rb/\mathrm{d}\ln t$ is the instantaneous expansion exponent, which takes the self-similar value $3/5$ only in the non-radiative limit (Sect.~\ref{sec:bubble_profiles}).
Consistency with the classical conduction-driven structure \citep{1977ApJ...218..377W} then requires the temperature and velocity profiles inside the bubble to satisfy
\begin{align}
    T &= \left(\frac{25}{4}\,\dot{M}_\mathrm{b}\,\frac{\kB}{4\pi \Rb^2\,\mu\,\mathcal{C}}\right)^{2/5}\left(1 - \zeta\right)^{2/5}\Rb^{2/5}, \label{eq:Tprofile} \\[4pt]
    \frac{\partial T}{\partial r} &= -\frac{2}{5}\frac{T}{\left(1 - \zeta\right)\Rb}, \label{eq:dTdr} \\[4pt]
    v &= \frac{\alpha \Rb}{t} - \dot{M}_\mathrm{b}\,\frac{\kB T}{4\pi \Rb^2\,\mu\,\Pb}, \label{eq:vprofile}
\end{align}
where $\mathcal{C} = 6\times10^{-7}~\mathrm{erg\,s^{-1}\,cm^{-1}\,K^{-7/2}}$ is the Spitzer conduction coefficient.

\subsection{Force budget on the shell}
\label{sec:EoM}

The radiation force combines direct photon absorption with infrared re-emission trapping:
\begin{equation}
    \Frad = \frac{\fabs^{\mathrm{bol}}\,L_\mathrm{bol}}{c}\left(1 + \tau_\mathrm{IR}\right),
    \label{eq:Frad}
\end{equation}
where $\fabs^{\mathrm{bol}}$ is the luminosity-weighted bolometric absorption fraction, $L_\mathrm{bol}$ is the bolometric cluster luminosity, and $\tau_\mathrm{IR}$ is the shell infrared optical depth \citep{2009ApJ...703.1352K,2010ApJ...709..191M,2017MNRAS.470.4453R}. Gravity combines the cluster potential with shell self-gravity:
\begin{equation}
    \Fgrav = \frac{G\,\Msh}{\Rb^2}\left(\Mstar + \frac{\Msh}{2}\right).
    \label{eq:Fgrav}
\end{equation}
The enclosed mass in Eq.~\eqref{eq:Fgrav} omits the pre-existing field stellar population, an approximation valid at natal-cloud scales but not for $\Rb \gtrsim 50\text{--}100\,\mathrm{pc}$ (Sect.~\ref{sec:caveat_grav-field}).

The driving pressure $\Pdrive$ in Eq.~\eqref{eq:motion} may receive contributions from three physical terms: 
the thermal pressure of the shocked wind bubble, $\Pb$; 
the thermal pressure of the photoionised gas, $\PHII$; and 
the ram pressure of newly injected wind and SN ejecta, $\Pram$.
These terms cannot be added blindly. 
The relative importance of the wind bubble and the photoionised region depends on cluster age and environment, because the wind bubble can either overpressure the ionised gas or remain subdominant to it \citep{2025ApJ...989...42L}. 
In the overpressured limit, the ionised gas collapses into a geometrically thin layer with $\Rb \simeq \Rif$. Pressure balance at the contact discontinuity then implies $\PHII \simeq \Pb$, so adding both pressures would count the same support twice \citep{2017MNRAS.470.4453R,2014ApJ...785..164M,2021MNRAS.501.1352G}. 
On the other hand, during the energy-driven phase, winds thermalise at the reverse shock $\Rts$, and their kinetic energy enters $\Eb$. Adding $\Pram$ at $\Rb$ would count the mechanical input once as stored thermal pressure and again as instantaneous momentum flux.

We avoided both double counts by selecting the driving term according to the dynamical phase.
In the energy-driven phase $\Pb$ already contains the thermalised wind contribution, so $\Pram$ is excluded. The photoionised layer is thin and in pressure balance with the bubble, such that $\PHII$ is bounded above by $\Pb$:
\begin{equation}
    \Pdrive^{\mathrm{(E)}} = \max\!\left(\Pb,\;\PHII\right).
    \label{eq:Pdrive_E}
\end{equation}
In the transition phase (see Sect.~\ref{sec:transition}), $\Pb$ declines as the hot bubble cools and freshly injected material begins to act more directly on the shell. We therefore compared the remaining bubble pressure with the additive momentum-phase support:
\begin{equation}
    \Pdrive^{\mathrm{(T)}} = \max\!\left(\Pb,\;\PHII + \Pram\right).
    \label{eq:Pdrive_T}
\end{equation}
In the momentum-driven phase, $\Eb \to 0$ and $\Pb \to 0$ by construction, leaving
\begin{equation}
    \Pdrive^{\mathrm{(M)}} = \PHII + \Pram.
    \label{eq:Pdrive_M}
\end{equation}
The $\max$ operator in Eqs.~\eqref{eq:Pdrive_E} and \eqref{eq:Pdrive_T} therefore prevents $\Pb$ and $\PHII$ from acting as independent pressures when they describe the same contact-discontinuity balance, and it hands control to $\PHII+\Pram$ once the stored bubble energy no longer dominates.

The bubble pressure $\Pb$ is the thermal pressure of the hot shocked interior (Eq.~\ref{eq:Pb}). It encodes the integrated history of mechanical energy injection, because wind and SN ejecta thermalise at $\Rts$ and accumulate in $\Eb$ before cooling or doing work on the shell.

The photoionised gas pressure is the thermal pressure of the ionised layer,
\begin{equation}
    \PHII = \frac{\muH}{\mup}\,\nshifStr\,\kB\,\Tion\;,
    \label{eq:PHII}
\end{equation}
where $\Tion \simeq 10^4\,\mathrm{K}$. 
Observed and modelled \hii regions show typical values near $\Tion\sim8000\,\mathrm{K}$ at solar abundance and can exceed $1.2\times10^4\,\mathrm{K}$ below $0.1\,Z_\odot$ \citep{1983MNRAS.204...53S,2020MNRAS.492..915G, 2024ApJ...964...47B}. 
The ionisation-front radius ($\Rif$) comes from the shell-structure calculation (Sect.~\ref{sec:shell}). Reading the density directly from the shell solution at $\Rif$ would tie $\PHII$ too closely to $\Pb$, because the inner-edge density is itself set by the bubble-pressure boundary condition. We therefore estimated the density from recombination balance. For a uniform density in the ionised volume,
\begin{equation}
    \nshifStr = \left(\frac{3\,f_{\mathrm{abs}}^{\mathrm{gas}}\,\Qi}{4\pi\left(1+\ZHe\xHe\right)\,\alpha_B\left(\Rif^3-\Rb^3\right)}\right)^{\!1/2},
    \label{eq:nIF_Str}
\end{equation}
where $f_{\mathrm{abs}}^{\mathrm{gas}}$ is the fraction of LyC photons absorbed by gas (see Sects.~\ref{sec:budget_diagnostics} and~\ref{sec:caveat_phii}). We assumed case-B recombination with a coefficient $\alpha_B = 2.59\times10^{-13}~\mathrm{cm^3\,s^{-1}}$ at $10^4\,\mathrm{K}$ \citep{2006agna.book.....O}. 
The factor, $(1 + \ZHe\,\xHe)$, accounts for the contribution of ionised helium to the free-electron density.

The ram pressure is the instantaneous momentum flux of newly injected wind and SN ejecta at the shell:
\begin{equation}
    \Pram = \frac{\Fram}{4\pi \Rb^2}\;.
    \label{eq:Pram}
\end{equation}
Unlike $\Pb$, it represents current momentum deposition in the hot bubble.

The external confining pressure is expressed as
\begin{equation}
    \Pext =
    \begin{cases}
    \PISM + \frac{\muH}{\mup}\,n_\mathrm{ext}\,\kB\,\Tion, & \fesc > 0, \\[6pt]
    \PISM, & \text{otherwise,}
    \end{cases}
    \label{eq:Pext}
\end{equation}
where $\PISM$ is the ambient ISM pressure, and the second term represents the back-pressure from gas ($n_\mathrm{ext}$) outside the shell once ionising photons leak into the surrounding medium.

We set the neutral-ISM pressure floor to $\PISM=0$ to isolate the feedback-driven dynamics from environmental confinement. The photoionised back-pressure in Eq.~\eqref{eq:Pext} remains active once $\fesc>0$.
Observed and inferred confining pressures around \hii regions span several orders of magnitude across galactic discs, centres, low-metallicity dwarfs, and nearby Galactic star-forming regions \citep{2021MNRAS.508.5362B,2024A&A...685A..46R,2024A&A...689A.352K,2024ARA&A..62..369S,2025ApJ...982..140P,2026A&A...706A..95B}. 
The dependence of feedback dominance on $\PISM$ is examined in Paper~II.

\subsection{Bubble interior cooling}
\label{sec:cooling}
The total cooling luminosity of the bubble in Eq.~\eqref{eq:energy} sets the bubble's internal pressure, the timescale over which $\Eb$ decays, and the moment at which the bubble crosses into the momentum-driven phase:
\begin{equation}
    \Lcool = \int n_\mathrm{H}\, n_e\, \Lnet\, \mathrm{d}V \;,
    \label{eq:Lcool_integral}
\end{equation}
where $n_\mathrm{H}$ and $n_e$ are the hydrogen nuclei and electron number densities and the net cooling function $\Lnet \equiv \Lambda_{\mathrm{cool}}(T,\ldots) - \Lambda_{\mathrm{heat}}(T,\ldots) > 0$ for a gas parcel that loses net energy to radiation.

\subsubsection{CIE and non-CIE regime}
\label{sec:cooling-processes}

The dominant cooling mode for plasmas of solar metallicity within the temperature range of $10^4$--$10^7~\mathrm{K}$ is by metal line transitions. At high temperatures, the plasma can be assumed to be in collisional ionisation equilibrium (CIE; $T\gtrsim10^{5.5}~\mathrm{K}$). In this regime $\Lnet$ reduces to a one-parameter function of temperature (for fixed abundances), so a tabulated cooling curve interpolated in $T$ is enough.

Our choice for non-CIE-to-CIE transition temperature follows from simple timescale arguments. The timescale on which gas radiates thermal energy is $\tcool\sim \mathcal{U}(\mathrm{d}\mathcal{U}/\mathrm{d}t)^{-1}\propto T/\Lambda(T)$ at fixed density, and the cooling function $\Lambda(T)$ peaks near $10^{5\text{--}5.5}$~K owing to metal-line emission (from e.g.\ Fe and O; see \citealt{2009A&A...508..751S}). 
By contrast, the ion fractions relax approximately on the recombination time $t_\mathrm{rec}\sim [n_e\,\alpha_\mathrm{eff}(T)]^{-1}$, with $\alpha_\mathrm{eff}(T)\propto T^{-\eta}$, $0.5\lesssim\eta\lesssim1.5$ \citep{1996ApJS..103..467V} so that $t_\mathrm{rec}$ lengthens as $T$ rises. For temperatures below the peak in the cooling function, one therefore expects $t_\mathrm{rec} \gtrsim \tcool$. The gas cools faster than charge states can re-equilibrate, and the ionisation state lags the instantaneous temperature (non-CIE; \citealp[see][]{1993ApJS...88..253S, 2007ApJS..168..213G,2013MNRAS.434.1043O}). 
At temperatures above the peak, however, $\Lambda(T)$ declines and $\tcool$ grows rapidly; the inequality reverses; and CIE becomes a good approximation.

This simplification breaks down whenever a strong photoionising radiation field is present or when the gas undergoes rapid thermal or dynamical evolution. Both conditions are generically satisfied inside \hii regions and feedback-driven bubbles: the central cluster continuously irradiates the surrounding gas with a time-evolving ultraviolet spectrum, and the bubble interior can cool rapidly once $\Eb$ begins to drain. Under non-CIE conditions, the charge-state distribution is set by competition between photoionisation, collisional ionisation, and recombination, and $\Lnet$ depends explicitly on four parameters: the gas temperature $T$, the gas density $n$, the ionising photon flux $\phi_i$, and the cluster age $t_\mathrm{age}$ \citep{2019MNRAS.483.2547R}. The density enters because it modulates the recombination rate, and the cluster age matters because older clusters produce softer spectra that are less effective at ionising metals. At high temperatures ($T \gtrsim 10^{5.5}\,\mathrm{K}$) the collisionally ionised fractions dominate even in the presence of a strong radiation field, so non-CIE corrections are small.

\subsubsection{Bubble interior structure}
\label{sec:bubble_profiles}

We used the following quantities to characterise the bubble interior:
\begin{align}
\label{eq:selfsim_exponents}
\alpha \equiv \frac{\mathrm{d}\ln \Rb}{\mathrm{d}\ln t},\qquad
\beta  \equiv -\,\frac{\mathrm{d}\ln \Pb}{\mathrm{d}\ln t},\qquad
\delta \equiv \left(\frac{\partial \ln T}{\partial \ln t}\right)_{\!\zeta}\;,
\end{align}
where $\zeta \equiv r/\Rb$, so that $\delta$ is evaluated at fixed fractional
radius.
With these definitions, the equations of energy and mass conservation can be written as \citep[see][]{1977ApJ...218..377W}
\begin{equation}
\begin{split}
\frac{1}{\Pb r^2} \frac{\partial}{\partial r} \left( \mathcal{C} T^{5/2} r^2 \frac{\partial T}{\partial r} \right)
- \frac{5}{2} \left( v - \frac{\alpha r}{t} \right) \frac{1}{T} \frac{\partial T}{\partial r}
\\ - \frac{n_e n_{\mathrm{H}} \Lambda_{\mathrm{net}}}{\Pb}
= \frac{\beta + 2.5 \delta}{t}\;,
\end{split}
\label{eq:bubble_energy_structure}
\end{equation}
\begin{align}
\frac{1}{r^2} \frac{\partial}{\partial r} \left( r^2 v \right)
- \left( v - \frac{\alpha r}{t} \right) \frac{1}{T} \frac{\partial T}{\partial r}
= \frac{\beta + \delta}{t}\;.
\label{eq:bubble_mass_structure}
\end{align}

If radiative losses are neglected ($\Lambda_{\mathrm{net}} = 0$), these equations admit the self-similar solution of \citet{1977ApJ...218..377W}, for which the exponents take the constant values $\alpha = 3/5$, $\beta = 4/5$ and $\delta = -6/35$. Once cooling becomes important (i.e. when the bubble age exceeds $\sim10^5~\mathrm{yr}$ for our models) the solution is no longer self-similar and the three exponents must be determined numerically.

Solving Eqs.~\eqref{eq:bubble_energy_structure} and~\eqref{eq:bubble_mass_structure} for a trial pair $(\beta,\delta)$ yields the radial profiles $T(r)$, $v(r)$ and $n(r)$ across the bubble, and hence the local emissivity. 
At each time step, \trinity\ therefore determines $(\beta,\delta)$ by requiring the interior solution to be self-consistent. The bubble energy loss implied by the structure equations must agree with that implied by the pressure--energy relation, and the temperature evolution implied by $\delta$ must agree with the structure solution. 
This gives two residuals in two unknowns, which are driven to zero numerically at every step. 
The solution step is detailed in Appendix~\ref{app:betadelta}.

\subsection{Cloud density profiles}
\label{sec:dens_main}

\trinity\ supports two initial density profile families: a piecewise power law (which includes the uniform cloud as the $\alpharho=0$ limit) and a BE sphere \citep{1955ZA.....37..217E,1956MNRAS.116..351B,2001Natur.409..159A}. 
The power law is simple and flexible but has limited physical motivation, whereas the BE sphere provides an idealised, centrally concentrated, pressure-confined profile of the type inferred for dense cores.

\subsubsection{Power-law profile}
\label{sec:dens_PL}

A simple power-law density distribution is described as
\begin{equation}
\rho_\mathrm{cloud}(r) =
\begin{cases}
\rhocore, & r \leq \Rcore, \\[6pt]
\rhocore \left( \dfrac{r}{\Rcore} \right)^{\alpharho}, & \Rcore < r \leq \Rcloud, \\[6pt]
\rho_\mathrm{ISM}, & r > \Rcloud .
\end{cases}
\end{equation}
Here, $\Rcore$ is the core radius and $\rho_\mathrm{ISM}$ is the ambient ISM density. We allowed $-2\leq\alpharho\leq0$, with a focus on $\alpharho = 0$ and $-2$. The limiting case $\alpharho = 0$ corresponds to a uniform cloud in which $\rhocore$ is the constant density throughout. This profile does not represent a structured molecular cloud, but it isolates the feedback physics in the simplest possible geometry. The opposite limit, $\alpharho = -2$, reproduces the $\rho \propto r^{-2}$ envelope of the singular isothermal sphere \citep{1969MNRAS.145..271L,1977ApJ...214..488S} but retains a finite-density core for $r \le \Rcore$ and truncates the cloud at $\Rcloud$. The core removes the central singularity, while the outer truncation keeps the total mass finite. Observed core radii span a broad range, from $\sim0.01$--$1\,\pc$ depending on the sample, tracer, and decomposition method \citep[see e.g.][]{2015A&A...584A..91K,2019A&A...628A.110M}. The power-law family thus brackets a useful range of idealised centrally concentrated structures.

\subsubsection{Bonnor--Ebert profile}
\label{sec:dens_BE}

A BE sphere satisfies hydrostatic equilibrium, Poisson's equation, and the isothermal equation of state:
\begin{align}
\frac{\mathrm{d}P}{\mathrm{d}r} &= -\rho(r) \frac{G M(r)}{r^2}\;, \label{eq:hydrostatic} \\[10pt]
\frac{1}{r^2} \frac{\mathrm{d}}{\mathrm{d}r} \left( r^2 \frac{\mathrm{d}\Phi}{\mathrm{d}r} \right) &= 4 \pi G \rho(r)\;, \label{eq:poisson} \\[10pt]
P &= \rho\, c_\mathrm{s}^2\;. \label{eq:isothermal}
\end{align}
Here, $c_\mathrm{s} = \left(\kB T/\mu\right)^{1/2}$ is the isothermal sound speed.
Integrating Eq.~\eqref{eq:poisson} once gives $\mathrm{d}\Phi/\mathrm{d}r = G M(r)/r^2$, so that Eq.~\eqref{eq:hydrostatic} can be written as $\mathrm{d}P/\mathrm{d}r = -\rho\,\mathrm{d}\Phi/\mathrm{d}r$.
Combining this with Eq.~\eqref{eq:isothermal} gives $c_\mathrm{s}^2\,\mathrm{d}\ln\rho/\mathrm{d}r = -\,\mathrm{d}\Phi/\mathrm{d}r$, which we integrate outwards from the centre to express the density in terms of the potential alone:
\begin{equation}
    \rho(r) = \rhocore \exp\{-\psi\}\;,\qquad
    \psi \equiv \frac{\Phi(r) - \Phi(0)}{c_\mathrm{s}^2}\;,
    \label{eq:BE_density_potential}
\end{equation}
where $\rhocore$ is the central mass density and the zero point of the dimensionless potential ($\psi$) is fixed by $\psi(0) = 0$.
Substituting Eq.~\eqref{eq:BE_density_potential} back into Poisson's Eq.~\eqref{eq:poisson} and introducing the non-dimensional radial parameter $\xi = \left( 4\pi G \rhocore / c_\mathrm{s}^2 \right)^{1/2} r$ yields a variant of the Lane--Emden equation \citep[see][]{ 1907gask.book.....E, 1939isss.book.....C}:
\begin{equation}
\frac{1}{\xi^2} \frac{\mathrm{d}}{\mathrm{d}\xi} \left( \xi^2 \frac{\mathrm{d}\psi}{\mathrm{d}\xi} \right) = e^{-\psi}\;.
\label{eq:lane-emden-isothermal}
\end{equation}
This equation has no closed-form solution and is integrated numerically subject to the boundary conditions,
\begin{align}
\psi(0) = 0\;, \quad \left. \frac{\mathrm{d}\psi}{\mathrm{d}\xi} \right|_{\xi = 0} = 0\;,
\label{eq:boundary-conditions}
\end{align}
the second of which follows from the requirement that the enclosed mass vanish at the centre.
The density profile then follows from Eq.~\eqref{eq:BE_density_potential}.

The solution is truncated at a dimensionless boundary radius $\xi_\mathrm{cl}$, which sets the density contrast between centre and edge, $\Omega = \rhocore/\rho(\xi_\mathrm{cl}) = \exp\!\bigl\{\psi(\xi_\mathrm{cl})\bigr\}$. 
Stable configurations exist only for $\xi_\mathrm{cl} \le 6.45$, corresponding to $\Omega \le 14.04$; beyond this critical value the sphere is gravitationally unstable and must collapse \citep{1956MNRAS.116..351B}. 
We adopted $\xi_\mathrm{cl} = 6.45$ as our default, placing the cloud at the marginally stable state\footnote{There is, however, some leeway for departures from strict BE stability. In particular, modest oscillations around the critical state are expected, driven by additional physical effects, such as internal magnetic pressure or external pressure arising from turbulence, nearby stellar winds, and radiation, which can act as stabilising mechanisms beyond purely thermal support.}.

\subsubsection{Shell mass and mass accretion rate}
\label{sec:dens_mass}
 
As the feedback-driven shell expands at velocity $\dot{R} = \mathrm{d}R/\mathrm{d}t$, the total swept-up mass within radius $R$ is
\begin{equation}
\Msh(R) = \int_0^{R} 4\pi r^2\,\rho_\mathrm{cloud}(r)\,\mathrm{d}r\,,
\label{eq:enclosed_mass}
\end{equation}
and the mass accretion rate follows as
\begin{equation}
\dot{M}_\mathrm{sw} = \frac{\mathrm{d}\Msh}{\mathrm{d}R}\,\dot{R} = 4\pi R^2\,\rho_\mathrm{cloud}(R)\,\dot{R}\,.
\label{eq:mdot}
\end{equation}
 
For the piecewise power-law profile, the enclosed mass for $\Rcore < R \le \Rcloud$ admits the closed form
\begin{equation}
\Msh(R) = 4\pi\rhocore\left[\frac{\Rcore^3}{3} + \frac{R^{3+\alpharho} - \Rcore^{3+\alpharho}}{(3+\alpharho)\,\Rcore^{\alpharho}}\right].
\label{eq:mass_powerlaw}
\end{equation}
 
For the BE sphere, the enclosed mass follows analytically from the Lane--Emden solution. Defining the dimensionless mass function $m(\xi) = \xi^2\,\mathrm{d}\psi/\mathrm{d}\xi$ \citep{1956MNRAS.116..351B}, the mass within radius $R$ is
\begin{equation}
\Msh(R) = \Mgas\,\frac{m(\xi)}{m(\xi_\mathrm{cl})}\,,
\label{eq:mass_BE}
\end{equation}
where $\xi = \xi_\mathrm{cl}\,(R/\Rcloud)$ and $m(\xi_\mathrm{cl})$ is the total dimensionless mass at the cloud boundary. The mass accretion rate is then
\begin{equation}
\dot{M}_\mathrm{sw} = \frac{\Mgas}{m(\xi_\mathrm{cl})}\,\frac{\mathrm{d}m}{\mathrm{d}\xi}\,\frac{\mathrm{d}\xi}{\mathrm{d}R}\,\dot{R} = 4\pi R^2\,\rhocore\,e^{-\psi(\xi)}\,\dot{R}\;,
\label{eq:Mdot_BE}
\end{equation}
where we used $\mathrm{d}m/\mathrm{d}\xi = \xi^2 e^{-\psi}$ from Eq.~\eqref{eq:lane-emden-isothermal} and $\mathrm{d}\xi/\mathrm{d}R = \xi_\mathrm{cl}/\Rcloud$. This recovers Eq.~\eqref{eq:mdot} exactly within the cloud boundary.

\subsubsection{Permitted parameter ranges}
\label{sec:dens_caveats}

Not every combination of $(\Mcloud,\,\ncore,\,\Rcore)$ produces a realisable cloud: the cloud radius must be plausible and the density profile must stay above the ambient ISM density at the boundary.
We discuss these constraints in Appendices~\ref{app:cloud_constraints} and \ref{app:density_smoothing}.

\subsection{Radiation-shell coupling}
\label{sec:shell}
The shell traps and reprocesses the radiation it absorbs, and the resulting density profile feeds back on $\fabs$ and $\tau_{\mathrm{IR}}$. The shell-structure system that closes this coupling is \citep[see also][]{2011piim.book.....D, 2014ApJ...785..164M}
\begin{align}
    \frac{d}{dr}\!\left(\frac{\muH}{\mup}\,n_{\rm sh}\,\kB\,\Tion \right)
      &= \frac{n_{\rm sh}\,\sigma_d}{4\pi r^2 c}\left(L_n e^{-\tau_d} + L_i \phi_{\rm f}\right) 
         \nonumber\\
      &\hspace{1em} + \frac{L_i}{\Qi\,c}\,\left(1+\ZHe\xHe\right)\,\alpha_B\,n_{\rm sh}^2
         \label{eq:shell_photonMomentum_ion}\;,\\
    \frac{d\phi_{\rm f}}{dr}
      &= -\frac{4\pi r^2}{\Qi}\,\left(1+\ZHe\xHe\right)\,\alpha_B\,n_{\rm sh}^2 \nonumber\\
    &\hspace{1em} 
         - n_{\rm sh}\,\sigma_d\,\phi_{\rm f}
         \label{eq:shell_absorptionRate_ion}\;,\\
    \frac{d\tau_d}{dr} &= n_{\rm sh}\,\sigma_d \;,
         \label{eq:shell_dustAttenuation_ion}
\end{align}
where $n_\mathrm{sh}(r)$ is the number density of hydrogen nuclei in the shell, $\Qi$ is the rate of ionising photons emitted by the central cluster, $\phi_{\mathrm f}(r)$ is the fraction of $\Qi$ that remains unabsorbed at radius $r$, $\Tion=10^4~\mathrm K$ is the ionised gas temperature, $L_i$, $L_n$ are the ionising and non-ionising luminosities, respectively, where $L_i+L_n=L_\mathrm{bol}$ is the bolometric luminosity. We set $\sigma_d = 1.5\times10^{-21}~\mathrm{cm}^2$ \citep{2011ApJ...732..100D, 2017MNRAS.470.4453R} as the dust absorption cross-section per hydrogen nucleus. We obtain the star cluster feedback parameters (i.e. $L_i, L_n, \Qi$) directly with \textsc{starburst99} \citep{1999ApJS..123....3L}.

Here, Eq.~\eqref{eq:shell_photonMomentum_ion} describes the momentum deposited in the shell gas by dust absorption and by photoionisation.
Equations~\eqref{eq:shell_absorptionRate_ion} and~\eqref{eq:shell_dustAttenuation_ion} describe radiation-gas coupling (recombination and dust absorption) and optical depth, respectively. 
We hold $\sigma_d$ constant (see Sect.~\ref{sec:caveat_shell-dust} for justification). However, dust opacity varies with environment: lower-metallicity systems carry less grain mass, and the effective cross-section drops as the cluster ages and hard photons fade. We absorb only the metallicity scaling, $\sigma_{\mathrm{eff}} = (Z/Z_\odot)\,\sigma_d$. At present, all \trinity\ models are run at solar metallicity ($Z=Z_\odot$); a full metallicity-dependent treatment is deferred to future work beyond the current paper series. 

We assumed that the central cluster evacuated a low-density cavity, so absorption inside this zone is negligible. Such an idealisation is common in modelling \hii regions \citep[see][]{2011piim.book.....D, 2014ApJ...785..164M}. 
Thus, we imposed the conditions $\phi_{\mathrm f}(\Rb) = 1$ and $\tau_d(\Rb) = 0 $.
A further boundary condition follows from pressure equilibrium across the contact discontinuity \citep{2017MNRAS.470.4453R}. The ionised gas at the shell base matches the bubble thermal pressure, $\Pb$, giving the minimum shell density, 
\begin{equation}
    \nshb = \left(\mup/\muH\right)\,\Pb/\left(\kB\,\Tion\right)\;.
    \label{eq:nShell0}
\end{equation}

We integrated outwards until the ionising budget was exhausted (i.e. $\phi_{\mathrm f}(r)\rightarrow0$).
Once the shell grows sufficiently thick, the ionisation front slows down; the outer layer therefore cools and forms a neutral layer that is dynamically coupled to the radiation field predominantly through dust absorption of the non-ionising continuum that leaks from the interior. In this regime, the equations above simplify to
\begin{align}
\frac{d}{dr} \left(\frac{\muH}{\mun}\,n_{\rm sh}\,\kB\,\Tneu \right)
   &= \frac{n_{\rm sh}\,\sigma_d}{4\pi r^2 c}\,L_n e^{-\tau_d}\;,\label{eq:shell_photonMomentum_neu}\\
\frac{d \tau_d}{dr} &= n_{\text{sh}} \sigma_d \label{eq:shell_dustAttenuation_neu}\;.
\end{align}
We adopted $\Tneu=100\,\mathrm{K}$ for the cold neutral layer.

\subsection{Energy-to-momentum transition}
\label{sec:transition}

The bubble transitions from energy- to momentum-driven once radiative losses drain its thermal reservoir. From Eq.~\eqref{eq:energy}, the bubble remains energy-driven while $\Lmech$ exceeds the combined drain from cooling losses ($\Lcool$) and $P\,\mathrm{d}V$ work on the shell. We triggered the transition when $\Lcool$ approached $\Lmech$, at which point net heating of the bubble became small compared to its radiative losses. We adopted $\Lcool/\Lmech>0.95$ as the numerical hand-off criterion \citep{1977ApJ...218..377W, 2009ApJ...693.1696H}. In the self-similar Weaver solution $\Lcool\propto t^{22/35}$ for $\Lambda\propto T^{-1/2}$, so cooling inevitably dominates at sufficiently high ambient density or late time. 

Once \trinity enters the transition phase, $\Eb$ decays at the sound-crossing rate,
\begin{equation}
    \frac{\mathrm{d}\Eb}{\mathrm{d}t} = -\frac{\Eb}{\tsc}, \qquad \tsc \equiv \frac{\Rb}{c_\mathrm{s}},
    \label{eq:Ebdecay}
\end{equation}
with $c_\mathrm{s}$ denoting the sound speed in the hot interior. The transition phase ends when the ram-pressure fraction $\Pram/(\Pb+\Pram)$ exceeds $0.9$, at which point the code switches to the momentum-driven equations.

As $\Eb\to 0$, wind and SN ejecta impinge directly on the shell as a momentum flux $\Pram$ instead of thermalising at the inner shock. 
The shell then coasts on the momentum accumulated during the energy-driven phase and decelerates under gravity, with subsequent winds and SNe adding momentum directly.

\subsection{Stopping criteria}
\label{sec:collapse-dissolve}

The bubble has two natural endpoints. In one, gravity wins and the shell re-collapses. In the other, the bubble continues to expand until it dissolves into the ambient ISM. We evolved each model until one of two stopping criteria was met. The first is a temporal cutoff covering the epoch over which winds and the earliest SNe dominate the energy and momentum budget. The second is a spatial cutoff at which the structure ceases to be a coherent shell and instead merges with its surroundings. To this end, we adopted fiducial thresholds of $t_\mathrm{max}=15\,\Myr$ and $R_\mathrm{max}=500\,\pc$.

A later version of the code will add fragmentation pathways that can both drain energy and break shell coherence: gravitational fragmentation of the swept shell, Rayleigh--Taylor growth at the decelerating contact, and venting through low-density channels. With these in place, the spatial cutoff above can be replaced by a physical breakdown criterion.

\section{Validation}
\label{sec:validation}

\begin{figure}
  \centering
  \includegraphics[width=\columnwidth]{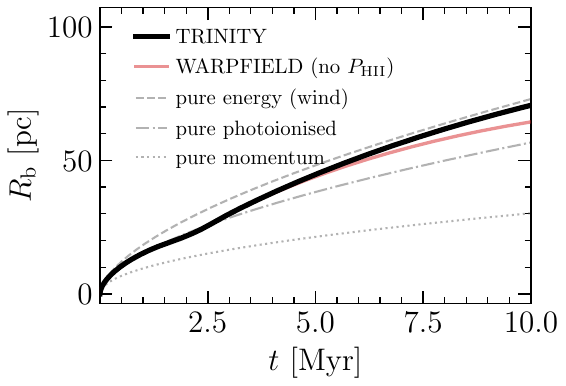}
    \caption{Evolution of the bubble radius for the fiducial test cloud ($\Mcloud = 10^6\,\Msun$, $\ncore = 10^3\,\ccm$, $\varepsilon = 0.01$), comparing \trinity\ (full physics) with a $\PHII$-disabled run of \trinity, as well as with three analytic limiting solutions in a uniform medium without gravity: the energy-driven Weaver wind, the photoionised Spitzer expansion, and the momentum-driven snowplow. 
    For this run, including $\PHII$ boosts the bubble radius at late times by $\sim 10\%$ (see Sect.~\ref{sec:validation}).}
  \label{fig:radiusComparison_sfe001_n1e3}
\end{figure}

We validated \trinity\ by overplotting its predicted radius trajectory $\Rb(t)$ on three analytic limiting solutions, and on a $\PHII$-disabled run of \trinity.
The three reference curves are the energy-driven wind solution $R \propto (L_{\mathrm{w}}/\rho_0)^{1/5}\,t^{3/5}$ \citep{1975ApJ...200L.107C, 1977ApJ...218..377W}; the photoionised expansion of \citet{1978ppim.book.....S}, with $R \propto t^{4/7}$ once the \hii region has overrun its initial Str\"omgren radius; and the momentum-driven snowplow $R \propto t^{1/2}$ in which the wind delivers momentum at a constant rate to a thin shell without thermal support.
Each scaling relation is a feedback-isolated solution in a uniform medium without gravity. 
They are not bounds on the \trinity\ trajectory, but scaling references for it.

Figure~\ref{fig:radiusComparison_sfe001_n1e3} shows the comparison for our fiducial test cloud, with $\Mcloud = 10^6\,\Msun$, $\ncore = 10^3\,\ccm$, and $\varepsilon = 0.01$. 
The cluster mass is $M_\star = \varepsilon\,\Mcloud = 10^4\,\Msun$, low enough that radiation pressure stays subdominant throughout and the comparison reduces to a clean test of the wind, ionised-gas, and momentum channels. 
\trinity\ tracks the pure-wind Weaver curve over the energy-driven phase, recovering the adiabatic wind limit at early times before cooling has integrated up to a significant fraction of the wind energy.
The trajectory then deviates from Weaver and rises more shallowly.
The late-time \trinity\ curve sits between the pure-photoionised and pure-energy reference solutions, showing that ionised-gas pressure and wind ram pressure both drive the shell.
The curve does not converge onto either limiting solution, since neither channel dominates on its own at this efficiency.

The $\PHII$-disabled run also quantifies the photoionisation contribution directly. By $t = 10\,\mathrm{Myr}$ the \trinity\ bubble radius is $\sim 10\%$ larger than the $\PHII$-disabled run, because $\PHII$ keeps the shell expanding while the cooled bubble in the comparison run loses its driving pressure. 
A complete mapping of the parameter space ($\Mcloud$, $\varepsilon$, $\ncore$, $\Rcore$) is the subject of the second paper of this series (Paper~II).

\section{First results}
\label{sec:results}

\subsection{Feedback budget and escape fractions}
\label{sec:budget_diagnostics}

\begin{figure}
    \centering
    \includegraphics[width=\columnwidth]{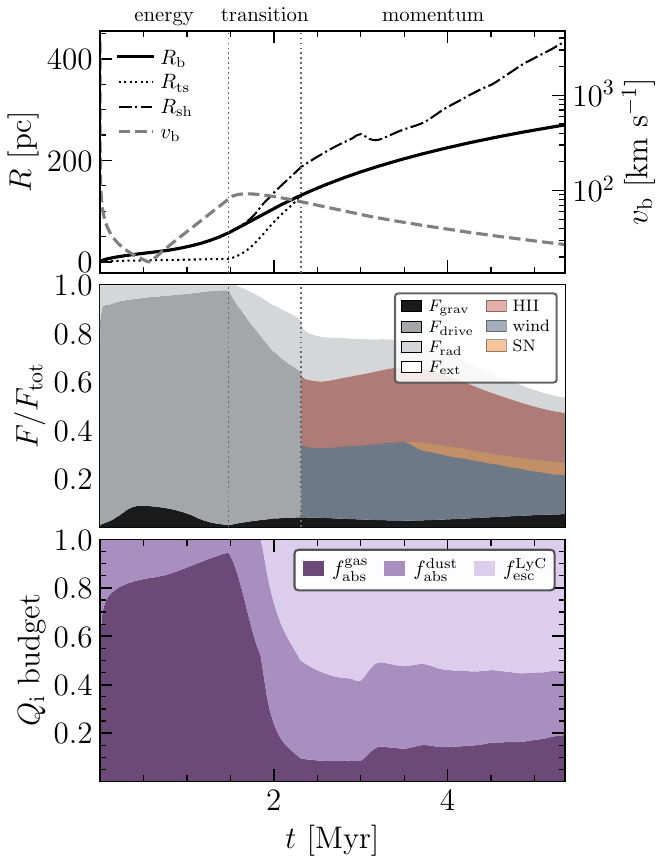}
    \caption{Diagnostic outputs of \trinity\ for a run with $M_{\mathrm{cloud}} = 10^6\,\Msun$, $\ncore = 10^3\,\ccm$, and $\varepsilon = 0.10$. 
    Top: Bubble radius $\Rb(t)$, reverse-shock radius $\Rts(t)$, shell radius $\Rshell(t)$, and bubble velocity $\vb(t)$, with the energy-driven, transition, and momentum-driven phases marked. 
    Middle: Fractional force budget on the shell, normalised to the sum of the force magnitudes. The outer stack marks the gravity ($F_{\mathrm{grav}}$), total driving force ($F_{\mathrm{drive}}$), radiation pressure ($F_{\mathrm{rad}}$), and external pressure ($F_{\mathrm{ext}}$). 
    Within $F_{\mathrm{drive}}$, the momentum phase is decomposed into wind, photoionised gas (\hii), and SN contributions. 
    Bottom: Fraction of ionising photons absorbed by gas $(f_{\mathrm{abs}}^{\mathrm{gas}}$), absorbed by dust ($f_{\mathrm{abs}}^{\mathrm{dust}}$), or escaping past the ionisation front ($\fesc$).}
    \label{fig:teaser_fiducial}
\end{figure}

With the dynamics validated against analytic limits at low cluster mass, we now demonstrate the full diagnostic suite at higher SFE. 
For the model in Fig.~\ref{fig:teaser_fiducial}, $\varepsilon=0.10$ gives $M_\star=10^5\,\Msun$, large enough that every feedback term is measurable. 
The figure reports three outputs from the same run: the shell trajectory, the force budget, and the ionising-photon budget. 

The top panel shows the dynamical trajectory.
The energy-driven phase lasts $\sim1.5\,\Myr$, after which hot-bubble cooling becomes efficient and the solution enters the momentum-dominated evolution. 
The middle panel shows the forces that act on the shell. We plot the fractional contributions to the sum of force magnitudes, not a signed acceleration budget. Gravity and external pressure therefore appear alongside the outward-driving terms even though they oppose expansion.

We kept $F_{\mathrm{drive}}$ as a single band during the energy-driven and transition phases and decomposed it into wind, photoionised gas (\hii), and SN channels only in the momentum phase. 
This follows the driving prescription. During the energy-driven phase, $P_{\mathrm{drive}}=\max(\Pb,\PHII)$ assigns the full driving pressure to whichever thermal component is larger, so a channel split would flicker between bubble and \hii labels as the selected pressure changes. 
Once the shell enters the momentum phase and $P_{\mathrm{drive}}=\PHII+\Pram$, the decomposition becomes additive and well defined. In this phase, \hii retains a measurable share of the driving force, so photoionised gas pressure continues to support expansion alongside wind ram pressure and SN ejecta.

The radiation force remains subdominant throughout this run, consistent with analytic expectations that radiation pressure becomes dynamically important mainly in compact, high-column, or high-luminosity regions \citep[e.g.][]{2009ApJ...703.1352K, 2025ApJ...982..140P}. 
The external pressure term acts against expansion and grows once ionising photons escape the shell, because the surrounding gas becomes photoionised and exerts a back-pressure. 
We set $\PISM=0$ in this fiducial run to isolate the stellar-feedback terms from environmental confinement. This is a controlled simplification: thermal pressures in the neutral ISM span from solar-neighbourhood values of the order of $P/k_B\sim3\times10^3\,\mathrm{K\,cm^{-3}}$ to high-pressure tails above $10^5\,\mathrm{K\,cm^{-3}}$ in dynamically active environments \citep{2003ApJ...587..278W,2011ApJ...734...65J}.

The bottom panel tracks the fate of the cluster's ionising photons. The shell-structure calculation in Sect.~\ref{sec:shell} partitions the LyC photon rate into absorption by hydrogen, absorption by dust, and escape past the ionisation front. We define
\begin{align}
    f_{\mathrm{abs}}^{\mathrm{gas}}  &= \frac{4\pi\left(1+\ZHe\xHe\right)\,\alpha_B}{\Qi} \int_{\Rb}^{\Rif} n_{\mathrm{sh}}^2(r)\, r^2 \, dr,\label{eq:fgas_LyC}\\
    f_{\mathrm{abs}}^{\mathrm{dust}} &= \sigma_d \int_{\Rb}^{\Rif} n_{\mathrm{sh}}(r)\, \phi_{\mathrm f}(r)\, dr,\label{eq:fdust_LyC}\\
    \fesc  &\equiv \phi_{\mathrm f}(\Rif),\label{eq:fesc_LyC}
\end{align}
where $\phi_{\mathrm f}(r)$ is the surviving LyC photon fraction, normalised to $\phi_{\mathrm f}(\Rb)=1$, and the gas term assumes case-B recombination in fully ionised hydrogen. These terms satisfy
\begin{equation}
    f_{\mathrm{abs}}^{\mathrm{gas}} + f_{\mathrm{abs}}^{\mathrm{dust}} + \fesc = 1
    \label{eq:Qi_budget}
\end{equation}
by construction. At early times, the shell is ionisation-bounded; LyC photons are absorbed inside the shell, a neutral layer remains beyond the ionisation front, and $\fesc=0$. The gas channel takes most of the budget, while the dust share is set by the dust optical depth across the ionised layer. Once the shell becomes density-bounded, $\fesc$ rises and ionising photons leak into the diffuse ISM. 
This connects the cloud-scale calculation to the galaxy-scale ionising-photon budget. Nearby measurements already show that $\fesc$ varies strongly between regions, from 30~Doradus in the Large Magellanic Cloud \citep{2013A&A...558A.134D}, to \hii regions in nearby galaxies \citep{2023MNRAS.524.1191T, 2026A&A...708A.202S}. 
By tracking $\fesc$ on the same time grid as the shell dynamics and force budget, \trinity\ separates two related timescales: the time at which feedback changes the shell dynamics, and the time at which LyC photons begin to escape. This makes the model useful for subgrid prescriptions in galaxy-scale simulations, where ionising-photon escape often depends on feedback-cleared low-column-density channels below the resolved scale \citep{2014ApJ...788..121K}.

\subsection{Dependence on the cloud density profile}
\label{sec:dependence-density-profile}
 
\begin{figure}
    \centering
    \includegraphics[width=\linewidth]{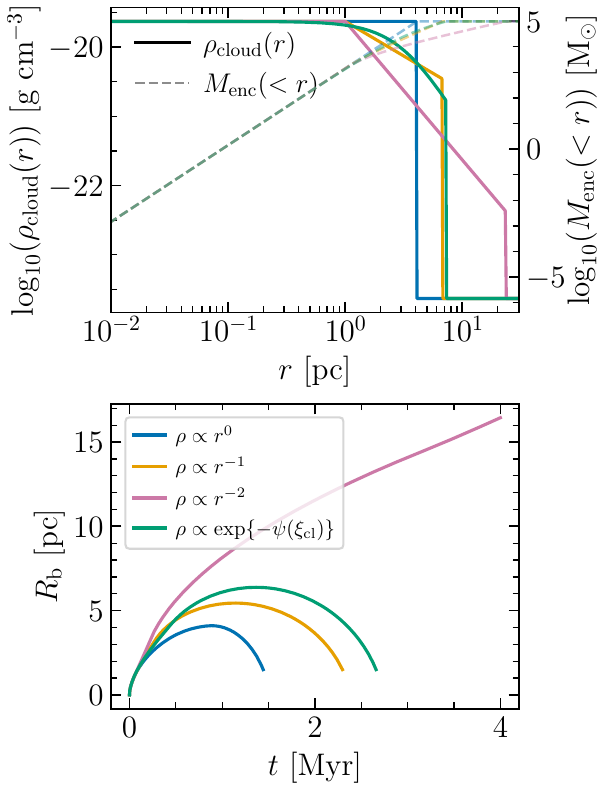}
    \caption{\textit{Top:} Density profiles (solid lines, left axis) and enclosed mass (dashed, right axis) for the four configurations in Sect.~\ref{sec:dependence-density-profile}. All share $\ncore = 10^4\,\ccm$, $\Mcloud = 10^5\,\Msun$, and $\varepsilon=0.01$. The steeper slopes concentrate mass towards the core, reducing the mass encountered by the expanding shell at large radii. 
    \textit{Bottom:} Bubble radius $\Rb(t)$. The bubble in the $\alpha_\rho = -2$ profile expands past $15\,\pc$ without turning around within the runtime, although it decelerates steadily after leaving the core. The remaining three profiles reach a maximum radius and experience re-collapse.}
    \label{fig:density_dependence}
\end{figure}

\begin{figure}
    \centering
    \includegraphics[width=\linewidth]{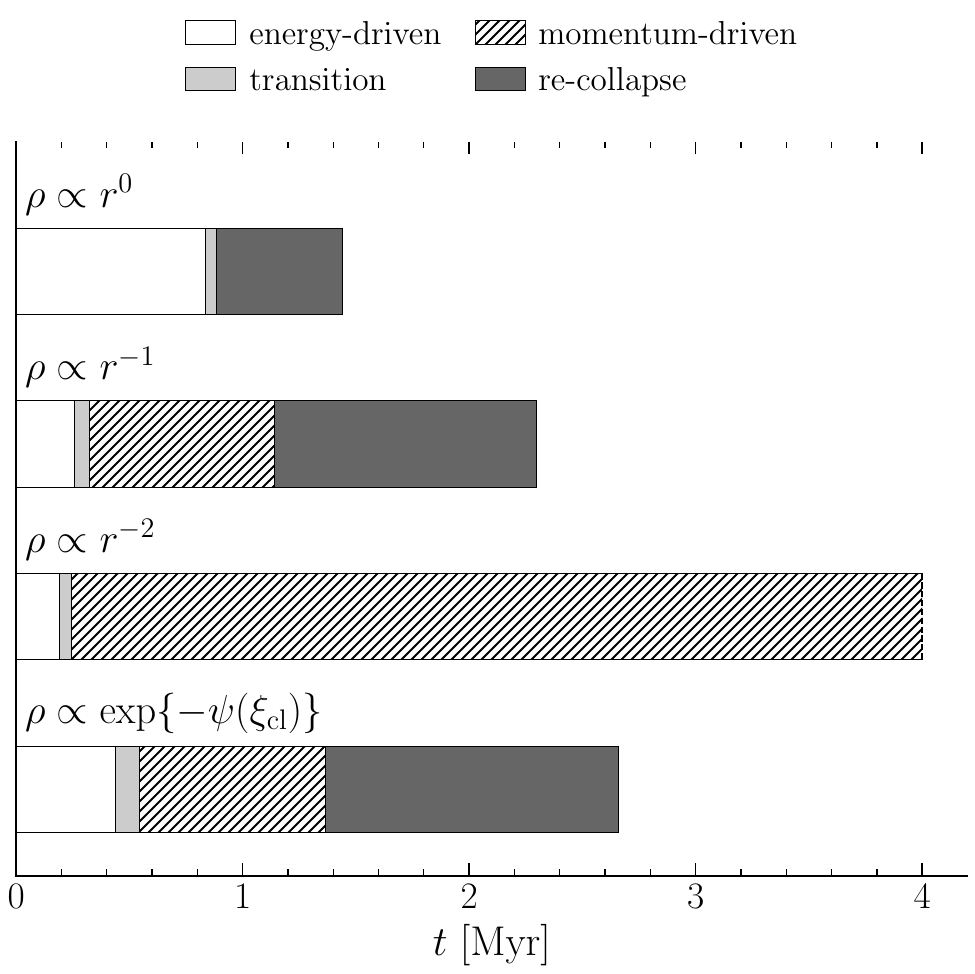}
    \caption{Phase timeline for the four density profiles in Sect.~\ref{sec:dependence-density-profile}. Each horizontal bar shows the duration of the energy-driven (white), transition (light grey), momentum-driven (hatched), and re-collapse (dark grey) phases. At fixed $\Mcloud$, $\ncore$, $\Rcore$, and $\varepsilon$, varying only the density profile shifts the phase durations and the final outcome. Only the $\alpha_\rho=-2$ cloud avoids re-collapse within the $4\,\Myr$ runtime.}
    \label{fig:phaseTime}
\end{figure}

To isolate the effect of cloud structure, we held cluster mass, cloud mass, and metallicity fixed and varied only the radial density profile. 
The fiducial cloud has $\Mcloud = 10^5\,\Msun$, $\ncore = 10^4\,\ccm$, core radius $\Rcore = 1\,\pc$, and $\varepsilon = 0.01$, corresponding to $\Mstar = 10^3\,\Msun$. 
The cloud is embedded in an ambient medium with $n_\mathrm{ISM}=1\,\ccm$, into which the shell may expand freely. 
We compared four profiles at fixed $\ncore$ and $\Mcloud$: power laws with $\rho\propto r^{\alpha_\rho}$ and $\alpha_\rho=0,-1,-2$, and a BE sphere truncated at $\xi_\mathrm{cl}=6.45$ (Sect.~\ref{sec:dens_BE}). 
Figures~\ref{fig:density_dependence} and \ref{fig:phaseTime} summarise the resulting shell histories.

At this low $\varepsilon$, the feedback budget sits near the dispersal threshold. The density profile therefore controls the shell trajectory by setting how fast the swept-up mass grows. While $\Rb<\Rcore$, all four clouds present the same central density to the bubble, so the early trajectories overlap in Fig.~\ref{fig:density_dependence}. Once the shell leaves the core, the runs diverge. In the homogeneous cloud, the shell encounters a mass reservoir that grows rapidly with radius, \(M_{\rm sh}\propto R^3\). The bubble remains compact, the shell loads mass efficiently, and the energy-driven phase lasts $0.84\,\Myr$. In the steep $\alpha_\rho=-2$ cloud, the density falls fast enough that the shell mass grows only linearly with radius outside the core. 
The shell therefore loads much less mass at large radii, decelerates far more slowly than in the uniform cloud, and leaves the energy-driven phase after $0.19\,\Myr$.

The BE sphere and the $\alpha_\rho=-1$ power law fall between these limits, with energy-phase durations of $0.44\,\Myr$ and $0.26\,\Myr$, respectively. Three profiles turn around and re-collapse on timescales of $t_\mathrm{coll}=1.4$--$2.7\,\Myr$. Only the $\alpha_\rho=-2$ cloud keeps expanding throughout the $4\,\Myr$ runtime\footnote{The $4\,\Myr$ endpoint is the runtime of this comparison. The shell may still decelerate and turn around at later times.}. In the steep profile, the shell reaches the low-density envelope before gravity and inertia can trap it. A steeper density profile therefore lowers the minimum $\varepsilon$ required for dispersal at fixed $\Mcloud$, $\ncore$, and $\Rcore$.

In this marginal case, the cloud outcome is decided before SNe dominate. The first core-collapse SNe occur after $\sim3$--$4\,\Myr$, by which time the re-collapsing models have already turned around and the expanding model has reached the low-density envelope. Thus, winds, radiation pressure, and photoionised gas pressure set the fate of the cloud in our study. Extragalactic measurements of cloud-scale efficiencies average over a distribution of density profiles, so population predictions such as bubble size distributions \citep[see e.g.][]{2023ApJ...944L..24W} should marginalise over that distribution rather than assume a canonical homogeneous cloud or a single power-law slope.

\subsection{On the emergence timescale of clusters}
\label{sec:emergence}

\begin{figure}
    \centering
    \includegraphics[width=\columnwidth]{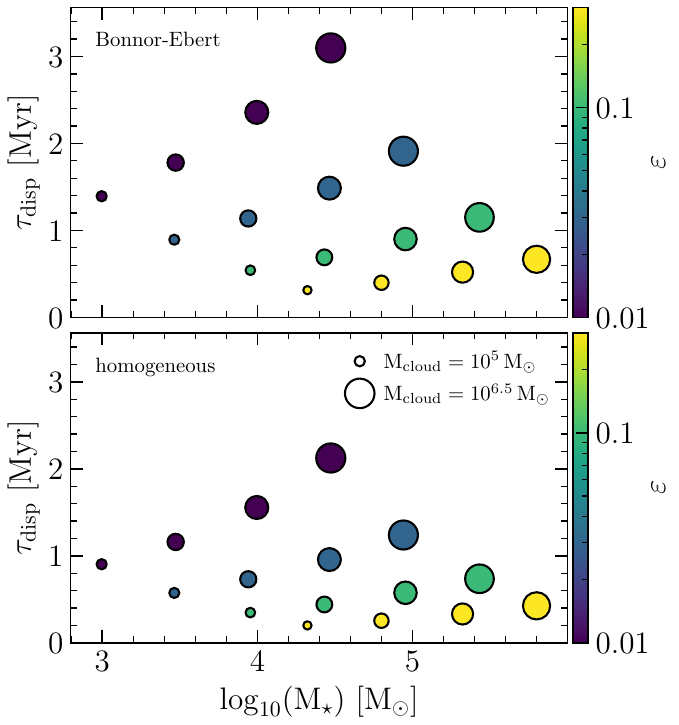}
    \caption{Dispersal timescale ($\tau_{\rm disp}$) as a function of cluster stellar mass ($M_\star$) at a fixed central density $\ncore = 10^3\,\ccm$. 
    \textit{Top:} BE profile. \textit{Bottom:} Homogeneous profile. 
    The marker colour encodes the SFE $\varepsilon \in [0.01, 0.3]$; and the marker size scales with $\Mcloud$ from $10^5$ to $10^{6.5}\,\Msun$. 
    At fixed $(\Mcloud,\varepsilon)$, the BE clouds disperse $\sim55\%$ later than the homogeneous clouds. This systematic offset shows that the assumed density profile changes the inferred emergence time even when dispersal is guaranteed. 
    Within each panel, increasing $\varepsilon$ shortens $\tau_{\rm disp}$ at fixed $\Mcloud$, while increasing $\Mcloud$ lengthens $\tau_{\rm disp}$ at fixed $\varepsilon$. These opposing trends weaken the apparent global relation with $M_\star$.}
    \label{fig:tau_disp_grid}
\end{figure}

To study the emergence timescales inferred from embedded and optically exposed cluster populations in nearby galaxies, we defined the dispersal timescale ($\tau_{\rm disp}$) as the time from cluster birth at which the bubble first reaches the natal cloud edge:
\begin{equation}
R_b(\tau_{\rm disp}) = R_{\rm cloud}\;.
\label{eq:tau_disp}
\end{equation}

Figure~\ref{fig:tau_disp_grid} compares $\tau_{\rm disp}$ across a $4\times4$ grid in $(\Mcloud,\varepsilon)$ at fixed central density $\ncore = 10^3\,\ccm$, with $M_\star=\varepsilon\Mcloud$. 
Across our sixteen matched $(\Mcloud,\varepsilon)$ pairs, the BE cloud disperses $55.1\pm2.8\%$ later than the homogeneous cloud.
The two models contain the same mass and central density, but the BE sphere spreads it over a larger radius\footnote{At a fixed mass and central density, the critical BE sphere is a factor of $R_\mathrm{BE}/R_\mathrm{hom}=1.79$ larger in radius. Since $\tau_{\rm disp}$ is set by $R_b=R_\mathrm{cloud}$, the geometric argument alone would give a $\sim79\%$ offset. However, an additional run matching both cloud mass and radius still demonstrates that the BE cloud disperses $24\pm9\%$ later. This is because the central concentration places most of the mass at small radii and slows the shell there. Equivalently, assuming a homogeneous profile for a cloud of a given radius and mass underestimates $\tau_{\rm disp}$ by $\sim19\%$.}.
Once feedback disperses every model in the grid, the profile no longer decides survivability, but it still shifts the timing.
A homogeneous initial condition would, therefore, underestimate the emergence time.

The dependence on cluster mass is subtler. At fixed $\Mcloud$, increasing $\varepsilon$ from $0.01$ to $0.3$ raises $M_\star$ and shortens $\tau_{\rm disp}$ by roughly a factor of five, because the feedback input per unit cloud mass rises. 
This trend has the same sign as observational emergence studies, where more massive clusters tend to clear their surroundings faster \citep{Pedrini2026}. 
At fixed $\varepsilon$, however, increasing $M_\star$ also means increasing $\Mcloud$ and hence $R_{\rm cloud}$; the shell must cross a larger system, so $\tau_{\rm disp}$ increases by a comparable factor. 

Across both density profiles, $\tau_{\rm disp}$ lies in the range $0.2$--$3.1\,\Myr$, less than the time required for the first core-collapse SN to occur. 
In this grid, the cloud outcome is therefore set by pre-SN feedback. 
This spans the megayear-scale window inferred from CO-to-H$\alpha$ spatial decorrelation measurements in nearby galaxies \citep{2020MNRAS.493.2872C,2019Natur.569..519K}, and is consistent with earlier \textsc{warpfield} calculations and with nearby-galaxy timing measurements that identify pre-SN feedback as the main driver of molecular-cloud dispersal \citep{2017MNRAS.470.4453R, 2026A&A...706A.186R}. 
Paper~II extends this comparison by varying $\ncore$ and imposing empirical cloud-density scaling, so that the simulated tracks can be compared directly with the mass-dependent emergence sequence.

\section{Limitations}
\label{sec:caveats}

\subsection{Geometry and initial conditions}
\label{sec:caveat_geometry}

Real regions do not evolve so cleanly. 
Density gradients drive blister and champagne flows \citep{1979A&A....71...59T,2020MNRAS.492..915G}. Moreover, turbulent clouds open low column density escape channels through which hot gas vents preferentially, rather than pushing isotropically on the whole shell \citep{2009ApJ...693.1696H,2024ApJ...970...18L}.
Our 1D treatment averages over this angular structure and replaces it with a single effective radius.

Spherical symmetry is a better approximation than it might at first appear. 
Many observed \hii regions are approximately round in projection: the Rosette Nebula \citep{1985A&A...144..171C}, the Orion Veil shell \citep{2019Natur.565..618P, Soler2026}, RCW~120 \citep{2021SciA....7.9511L}, and a wide variety of Galactic infrared ring nebulae \citep{2006ApJ...649..759C}. 
Extragalactic surveys show the same mixture of order and complexity. Most young, compact \hii regions in nearby spirals are well described by circular or mildly elliptical apertures \citep{2022MNRAS.512.1294H, 2026A&A...706A..95B}, though older superbubbles increasingly deviate from sphericity as galactic shear and bubble merging take over \citep{2023ApJ...944L..24W}. 
The Magellanic Clouds provide a nearby laboratory for this range of morphologies. MCELS and DeMCELS \citep{1999IAUS..190...28S,2024ApJ...974...70P} map the ionised gas across the LMC at parsec-scale resolution, from relatively simple shells and superbubbles to the highly structured 30~Doradus complex.
The same machinery applies in principle to compact star-forming systems at high redshift, where JWST resolves dense young complexes at $z \gtrsim 10$ \citep{2024Natur.632..513A}.
Extending the code to that regime requires adapting the input physics to low-metallicity and high-density conditions, including modified cooling and a top-heavy IMFs, all of which we defer to future work.

The cloud density profile $\rho(r)$ is set at $t=0$, and material beyond the expanding bubble does not evolve. It neither collapses, fragments, forms new stars, nor accretes onto the central cluster during the run.
This is a reasonable approximation while the shell is fast compared with the local free-fall time, which holds over most of the dispersing models here.
Observed GMCs are broadly consistent with turbulent or quasi-equilibrium support on cloud scales \citep{2009ApJ...699.1092H,2004RvMP...76..125M}, whereas dense sub-structures can be formally sub-virial \citep{2013ApJ...779..185K}. 
For these cases, coupling the shell dynamics to a live treatment of cloud self-gravity and turbulent evolution is left to future work.

\subsection{Gravity beyond the natal cloud}
\label{sec:caveat_grav-field}

Our gravity term in Eq.~\eqref{eq:Fgrav} includes only the young cluster and the swept-up shell. We neglected the pre-existing stellar component of the host galaxy. Disc stars older than the natal cloud whose enclosed mass adds to the inward force as the shell expands. 
On natal-cloud scales this omission is small, because the shell encloses little of the surrounding disc. It becomes less secure once $\Rb$ reaches tens of parsecs. 
For a local old-stellar surface density $\Sigma_\star$, the enclosed background stellar mass scales as $M_{\rm old}(\Rb)\sim \pi \Rb^2 \Sigma_\star$, so it can become comparable to or larger than $\Mstar$ at $\Rb\sim50$--$100\,\mathrm{pc}$ in normal galactic discs. 
Recent PHANGS measurements of $\sim18\,000$ H\,\textsc{ii} regions find that older stars dominate over young stars on $\gtrsim10\,\mathrm{pc}$ scales, and that self-gravity becomes competitive with pre-SN feedback on scales of order $60$--$130\,\mathrm{pc}$ depending on the environment \citep{2025ApJ...993L..20P}. 
Adding a radius-dependent term $M_{\rm old}(\Rb)$ from an axisymmetric stellar disc model is straightforward, but we defer this extension to the version of \trinity\ aimed at shell evolution beyond the parent cloud.

\subsection{Missing physics in the bubble interior}
\label{sec:caveat_bubble}
 
We computed bubble energy losses with the conductive-evaporation prescription following \citet{1977ApJ...218..377W}. The resulting cooling rate assumes classical electron heat conduction across a smooth, spherical contact discontinuity. Three-dimensional simulations and analytic models show that Kelvin--Helmholtz and Rayleigh--Taylor instabilities can shred the contact discontinuity into a corrugated, fractal-like surface; the enlarged interface enhances turbulent mixing and radiative cooling relative to a smooth spherical area \citep[e.g.][]{2021ApJ...914...89L,2024ApJ...970...18L}. A 1D code cannot resolve this geometry directly; it can only absorb the effect into an effective cooling prescription. In the absence of such a subgrid enhancement, \trinity likely keeps the hot bubble overpressured for too long. We therefore interpret the time at which our models reach the momentum-driven phase as an upper limit when turbulent mixing dominates the real interface.

In \trinity, three pieces of feedback physics are currently omitted from the dynamics. The first is magnetic support. 
At the cloud densities in our grid ($n \sim 10^{3}$--$10^{4}\,\mathrm{cm^{-3}}$), the upper-envelope field strengths compiled by \citet{2012ARA&A..50...29C} give magnetic pressures $P_{\mathrm{mag}}/\kB \sim 1$--$30\times10^{5}\,\mathrm{K\,cm^{-3}}$.
This pressure lies below the thermal pressure of photoionised gas for the fiducial models considered here, and below the early wind ram pressure in our model outputs. 
The approximation weakens at the upper end of the density grid, where  $P_{\mathrm{mag}}\propto n^{4/3}$ \citep{2012ARA&A..50...29C}.

Second, we included cosmic-ray ionisation in the cooling tables but neglected cosmic-ray pressure in the shell momentum equation. 
This is a reasonable approximation here, as the smooth Galactic cosmic-ray background does not exert a considerable net force on a spherical shell. Only a cosmic-ray pressure gradient across the bubble would contribute to the equation of motion.
Such gradients can matter on galactic scales and in SN-driven environments, but they are not part of the pre-SN, single-cloud models considered here \citep{2015ARA&A..53..199G,2012A&ARv..20...49V}.
Pre-SN acceleration at stellar-wind termination shocks remains possible, but its impact depends on efficient confinement and on the geometry; it is therefore unlikely to dominate the shell dynamics in the ordinary GMC regimes modelled here \citep[e.g.][]{2021MNRAS.504.6096M,2024MNRAS.527.3818B}.

We also neglected turbulent pressure inside the ionised gas. In the Orion Nebula, the non-thermal velocity dispersion is subsonic. \citet{2016MNRAS.463.2864A} find that $\sigma_\mathrm{turb} \approx 4$--$5\,\mathrm{km\,s^{-1}}$, below the ionised sound speed $c_i \approx 11\,\mathrm{km\,s^{-1}}$. Radiation-hydrodynamic simulations of \hii regions in turbulent molecular clouds reach a comparable subsonic state within a few sound-crossing times \citep{2014MNRAS.445.1797M}. This approximation should be safest for compact and intermediate \hii regions. It becomes weaker for giant regions such as 30\,Doradus, where observed line widths are supersonic and cluster winds, shells, and large-scale flows shape the gas kinematics \citep{2021A&A...649A.175M,2023MNRAS.523.4202G}.

\subsection{Dust and shell leakage}
\label{sec:caveat_shell-dust}

We scaled the dust cross-section linearly with metallicity and ignore both grain sublimation and the evolution of the grain-size distribution.
Sublimation does destroy grains within $r_{\mathrm{sub}}$ of the cluster, but only a negligible fraction of the total dust mass is lost in the regime modelled here \citep{2011piim.book.....D}. 
Our prescription is nevertheless a good approximation during the dense embedded phase, when the shell column density is high and the dust is well coupled to the gas. 
It overestimates the radiation-pressure boost factor $(1+\tau_{\mathrm IR})$ at late times in luminous, low-density environments, where cumulative grain depletion reduces $\tau_{\mathrm IR}$ below its initial value.

We also assumed a closed shell throughout (i.e. covering fraction $C_\mathrm{f} = 1$). 
The code does not evaluate fragmentation criteria or transition to a leaky-shell regime.
Once the shell opens, hot gas escapes, ionising photons leak out, and the shell captures less of the available momentum. \citet{2009ApJ...693.1696H} showed for Carina that the dynamics depend on the sky-averaged covering fraction rather than on the detailed shape of individual holes, which makes a parametrised $C_\mathrm{f}(t)$ a sensible next step.
Implementing an evolving covering fraction and propagating it into $\fesc$ will be deferred to future work.

\subsection{Treatment of photoionised gas pressure}
\label{sec:caveat_phii}

The crux of the problem is how photoionised gas pressure should enter the equation of motion.
Our construction in Sect.~\ref{sec:EoM} reduces, but does not remove, the coupling between $\PHII$ and $\Pb$. 
The inner-edge density depends on $\Pb$, the hydrostatic shell structure depends on that boundary value, and $\Rif$ enters Eq.~\eqref{eq:nIF_Str} through the resulting geometry. 
The recombination-balance estimate is therefore less directly tied to $\Pb$ than the local shell density at $\Rif$, but it is not an independent two-radius wind-bubble and \hii-region model. 
Evolving $\Rb$ and $\Rif$ as separate dynamical variables would provide that treatment; we leave this extension to future work.

Existing semi-analytic treatments make different architectural choices. 
One approach starts from a photoionisation-driven \hii-region solution and adds winds, radiation pressure, gravity, or photon breakout as correction terms to the same ionisation-front dynamics \citep{2020MNRAS.492..915G}. 
A different approach, followed by \citet{2025ApJ...989...42L}, carries separate radii for the wind-bubble boundary $\Rb$ and the ionisation front $\Rif$, then evolves the wind bubble and photoionised region as coupled parts of one feedback bubble through force balance and ionisation--recombination equilibrium. The third approach \citep{2017MNRAS.470.4453R,2019MNRAS.483.2547R, 2026A&A...709A.100J} sits closer to the single-radius class: $\PHII$ enters the hydrostatic shell structure through the inner boundary condition at the shell face, but does not appear as an explicit force in the shell equation of motion. 
\trinity lies between the two-radius and single-radius classes: it includes $\PHII$ explicitly in the shell equation of motion, but $\Rif$ is diagnosed from the shell structure rather than evolved as an independent dynamical variable.

\section{Summary and conclusions}
\label{sec:conclusions}

We presented \trinity, a 1D spherical thin-shell model that follows the expansion of feedback-driven bubbles in molecular clouds under stellar winds, SNe, direct and dust-reprocessed radiation pressure, photoionised gas pressure ($\PHII$), and gravity in a single, self-consistent, time-dependent calculation. \trinity\ succeeds \textsc{warpfield} \citep{2017MNRAS.470.4453R, 2019MNRAS.483.2547R} and introduces a phase-aware driving prescription (Sect.~\ref{sec:EoM}) together with flexible initial cloud structures (uniform, power-law, and BE).
We validated the code against the analytic Weaver wind and photoionisation limits and used a controlled set of experiments, including a fiducial low-efficiency cloud, a higher-efficiency example, and grids spanning cloud mass, central density, SFE, and density profile, to assess how each feedback channel and the cloud's internal structure shape its evolution. 
Our main conclusions are as follows.

\begin{enumerate}
    \item \textbf{The assumed internal cloud structure, in addition to the cloud mass and central density, sets whether and when a cloud disperses.}
    Of the four profiles tested at fixed $\Mcloud$ and $\ncore$ (uniform, $\alpha_\rho=-1$, $\alpha_\rho=-2$, and a truncated BE sphere), only the $\alpha_\rho=-2$ cloud keeps expanding within the $4\,\Myr$ runtime. The other three turn around and re-collapse on $1.4$--$2.7\,\Myr$ timescales (Sect.~\ref{sec:dependence-density-profile}).
    In the regime where every cloud does disperse, BE clouds reach a $\tau_{\rm disp}$ about $55\%$ later than the homogeneous clouds of the same mass and central density (Sect.~\ref{sec:emergence}).
    The minimum efficiency for cloud destruction and the inferred emergence time, therefore, depend on how mass is arranged inside the natal cloud.

    \item \textbf{Photoionised gas pressure is a dynamically important driver, particularly once the hot bubble has cooled.}
    Including $\PHII$ increases the bubble radius by $\sim10\%$ at $10\,\Myr$ relative to the $\PHII$-disabled run in the fiducial cloud. The separation grows after $\approx4\,\Myr$ as the cooled bubble loses thermal support (Sect.~\ref{sec:validation}).
    At higher efficiencies, photoionised gas pressure continues to contribute a measurable fraction of the driving even after the transition to momentum-dominated evolution.
    In the fiducial cloud, omitting $\PHII$ removes a driving term of the order of the ram pressure once the bubble has cooled and correspondingly underestimates the late-time expansion.

    \item \textbf{No single feedback channel dominates throughout, so feedback dominance cannot be inferred from the stellar population alone.}
    The energy-driven phase can be as short as $0.2\,\Myr$ in a steep density profile and lasts $\approx1.5\,\Myr$ in our higher-efficiency uniform cloud, before cooling pushes the solution into momentum-dominated evolution (Sect.~\ref{sec:budget_diagnostics}).
    In our higher-efficiency example, radiation pressure remains subdominant while winds, photoionised gas, and the direct ram pressure of freshly injected ejecta trade off over time.
    Which channel dominates is thus both time- and cloud-dependent and emerges only from evolving the channels together rather than assuming one a priori.

    \item \textsc{Trinity} \textbf{follows the shell dynamics, the force budget, and the ionising-photon budget together, and separates cloud dispersal from the escape of ionising radiation.}
    Because these are all computed on the same time grid, we can identify which feedback channel drives the shell at any given moment.
    The model also distinguishes two timescales that are easily conflated: the time at which the cloud is dispersed and the distinct time at which ionising photons begin to leak into the surrounding ISM.

\end{enumerate}

Whether a cloud is cleared, and how quickly, cannot be determined from its stellar population alone.
Internal density structure matters as much as stellar content, and the feedback channels must be followed together, since that which dominates changes with time and with the cloud.
\trinity\ provides the diagnostics needed to map the wind-, photoionisation-, radiation-, and SN-dominated regimes across cloud parameter space at a low cost and predicts dispersal and leakage timescales that can be directly compared with resolved observations.
Being 1D, \trinity\ averages over angular structure and does not capture champagne flows, blister breakouts, or the corrugated wind-bubble interface carved by hydrodynamic instabilities. Within this limitation it offers a fast, physically transparent description of feedback-driven bubbles.
In the second paper of this series (Paper~II), we will apply \trinity\ across the full range of parameter space and compare directly with resolved surveys.

\section*{Data availability}
The \trinity source code, input configurations, and post-processing scripts used to produce the figures in this paper are made publicly available at \url{https://github.com/JiaWeiTeh/trinity}. The figures can be regenerated by running \texttt{python paper/methods/make\_figures.py}.

\begin{acknowledgements}

We thank the anonymous referee for a careful and constructive report that improved the clarity of this paper.
We thank Moritz Sturm and Loke Ohlin for contributions to the code and for many helpful discussions during its development.
The team acknowledges financial support from the European Research Council (ERC) via Synergy Grant ``ECOGAL'' (project ID 855130) and from the German Excellence Strategy via the Heidelberg Cluster ``STRUCTURES'' (EXC 2181 - 390900948). In addition, RSK acknowledges funding from the German Ministry for Economy and Energy (BMWE) in project ``MAINN'' (funding ID 50OO2206), and from the German Science Foundation (DFG) as well as the French National Research Agency (ANR) for project ``STARCLUSTERS'' (funding ID KL 1358/22-1). RSK is grateful for computing resources provided by the Ministry of Science, Research and the Arts (MWK) of the State of Baden-W\"{u}rttemberg through bwHPC and DFG through grants INST 35/1134-1 FUGG and 35/1597-1 FUGG, and also for data storage at SDS@hd funded through grants INST 35/1314-1 FUGG and INST 35/1503-1 FUGG. KK gratefully acknowledges funding from DFG in the form of an Emmy Noether Research Group (grant number KR4598/2-1) and the ERC Starting Grant ``ISM-METALS'' (project ID 101077573).

\end{acknowledgements}

\bibliographystyle{aa}
\bibliography{ref}

\begin{appendix}

\section{Determination of the cooling parameters \texorpdfstring{$\beta$}{beta} and \texorpdfstring{$\delta$}{delta}}
\label{app:betadelta}

Once radiative loss begins, the equations defined in Sect.~\ref{sec:bubble_profiles} are no longer fixed at their self-similar values and must be determined at each timestep. 
Given the state $\{\Rts,\Rb,\Eb,\Pb\}$ and a trial pair $(\beta,\delta)$, the bubble energy loss rate can be computed in two independent ways.

The first, $\dot{E}_{\mathrm{b,E}}$, follows from the energy equation, Eq.~\eqref{eq:energy}: solving the structure equations~\eqref{eq:bubble_energy_structure} and~\eqref{eq:bubble_mass_structure} for the trial pair yields $T(r)$, $v(r)$ and $n(r)$, and hence $\Lcool$ through Eq.~\eqref{eq:Lcool_integral}.

The second, $\dot{E}_{\mathrm{b,PE}}$, follows from the definition of the bubble's thermal energy as seen in Eq.~\eqref{eq:Pb}. 
We write the rate of change of bubble thermal energy, and subsequently the rate of change of bubble volume as
\begin{align}
    \dot{\Eb} &= \frac{3}{2}\left(\dot{P}_{\rm b}\,\Vb + \Pb\,\dot{V}_{\rm b}\right)\;,\\
    \dot{\Vb} &= 4\pi\left(\Rb^{2}\vb - \Rts^{2}\dot{R}_{\rm ts}\right)\;.
    \label{eq:Ebdot_physical}
\end{align}
where $\dot{\Pb} = -\beta\Pb/t$ follows from the definition of $\beta$ in Eq.~\eqref{eq:selfsim_exponents}.
At $\Rts$ the ram pressure equals the bubble pressure, therefore differentiating $\Rts^{2} = \Fram/4\pi\Pb$ gives
\begin{equation}
    \dot{\Rts} = \frac{\Rts}{2}\left(\frac{\dot{F}_{\rm ram}}{\Fram} - \frac{\dot{P}_{\rm b}}{\Pb}\right)
               = \frac{\Rts}{2}\left(\frac{\dot{F}_{\rm ram}}{\Fram} + \frac{\beta}{t}\right)\;.
    \label{eq:Rtsdot}
\end{equation}
Substituting the above into Eq.~\eqref{eq:Ebdot_physical} one obtains a single explicit expression,
\begin{equation}
    \dot{\Eb}
      = -\,\frac{\beta\Eb}{t}
      \;+\; \frac{3\Eb}{\Rb^{3}-\Rts^{3}}\,\Rb^{2}\vb
      \;-\; \frac{3}{4}\,\Rts\!\left(\dot{F}_{\rm ram} + \frac{\beta\Fram}{t}\right)\;,
    \label{eq:Ebdot_PE}
\end{equation}
Thus, the pair $(\beta,\delta)$ is fixed by requiring the interior solution to be self-consistent in both energy and temperature:
\begin{align}
    \mathcal{R}_{E}(\beta,\delta) &= \dot{E}_{\mathrm{b,\,E}}(\beta,\delta) - \dot{E}_{\mathrm{b,\,PE}}(\beta,\delta)\;,\\
    \mathcal{R}_{T}(\beta,\delta) &= T(\beta,\delta) - T_{0}\;.
    \label{eq:beta_delta_residual}
\end{align}
where $T_0$ is the bubble temperature carried forward from the previous step. 
The solution is the pair for which $\mathcal{R}_{E} \simeq \mathcal{R}_{T} \simeq 0$.

In practice, \trinity\ minimises $\mathcal{R}_{E}^{2} + \mathcal{R}_{T}^{2}$ over $(\beta,\delta)$ within the physically motivated bounds $0\le\beta\le1$ and $-1\le\delta\le0$, for which the bubble pressure and temperature both decrease with time and the interior velocity remains outflowing\footnote{These bounds are conservative. In Paper~II we instead solved $\mathcal{R}_E=\mathcal{R}_T=0$ directly using an unbounded Newtonian-type root-finder on a better-conditioned residual. This approach reproduces the bounded solution where it converges, at a lower cost, and it follows brief feedback-driven re-pressurisation episodes in which $(\beta,\delta)$ leave the ranges above. During these episodes, the interior develops a weak, subsonic inward velocity that is dynamically negligible.}. The search starts from a coarse $5\times5$ grid centred on the previous time-step solution. Because the pair evolves smoothly between steps, the previous solution is an excellent initial guess and the grid search alone is usually sufficient.

\section{Cluster feedback input tables}
\label{app:stellar_feedback_table}

\trinity reads the cluster-integrated feedback budget from a single time-resolved table, computed in advance for a simple stellar population (SSP) cluster of mass $M_\star$ and metallicity $Z$. 
For each tabulated age ($t$), we require the following quantities: the ionising photon rate $Q_\mathrm{i}(t)$; the bolometric luminosity $L_\mathrm{bol}(t)$, split into ionising and non-ionising components $L_\mathrm{i}(t)$ and $L_\mathrm{n}(t)$ at $13.6\,\mathrm{eV}$; the wind and SN mechanical luminosities $L_\mathrm{w}(t)$ and $L_\mathrm{SN}(t)$; and the corresponding momentum injection rates $\dot{p}_\mathrm{w}(t)$ and $\dot{p}_\mathrm{SN}(t)$. 
For this paper, we generate the default tables with \textsc{starburst99} \citep[SB99;][]{1999ApJS..123....3L,2014ApJS..212...14L}, run with a \citet{2001MNRAS.322..231K} IMF over $0.1$--$100\,\Msun$ and Geneva tracks at the requested metallicity and rotation regime.

Nevertheless, \trinity\ is not tied to a specific stellar-population synthesis code. 
So long as atmospheric models are taken into account \citep[e.g. for $Q_\mathrm{i}$, $\dot M v_\infty$; see][]{1998ApJ...496..407H, 2001A&A...375..161P,2002MNRAS.337.1309S}, any combination of synthesis code and atmospheric models that delivers the eight quantities above on a common time grid can be substituted, including BPASS \citep{2017PASA...34...58E,2018MNRAS.479...75S}, which adds binary mass transfer, and SLUG \citep{2012ApJ...745..145D,2015MNRAS.452.1447K}, which samples the initial mass function (IMF) stochastically.

\section{Validity of the ZAMS assumption}
\label{app:ZAMS}

\begin{figure*}
\centering
\includegraphics[width=\textwidth]{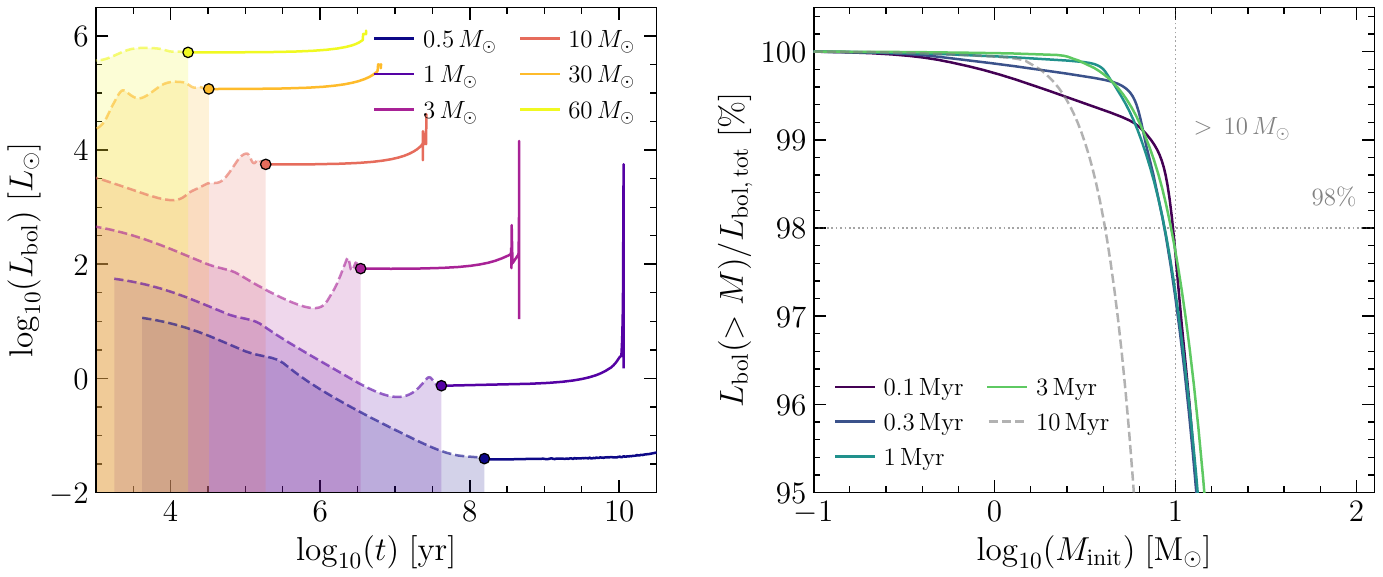}
\caption{\textit{Left:} MIST single-star evolutionary tracks at solar metallicity and zero rotation for the six initial masses between $0.5$ and $60\,\Msun$. 
PMS segments are illustrated with dashed lines, and the regions below them are shaded. The main-sequence and post-main-sequence segments are represented with solid lines.
The filled circles mark the zero-age main sequence. 
Massive stars ($M_{\mathrm{init}} \gtrsim 10\,\Msun$) reach the main sequence within $\lesssim 0.3\,\Myr$. 
\textit{Right:} Cumulative luminosity fraction $F(>M; t)$ from Eq.~\eqref{eq:Fgt}, evaluated for a $10^{6}\,\Msun$ \citet{2001MNRAS.322..231K} cluster at five ages drawn from MIST basic isochrones. 
Curves at $t \leq 3\,\Myr$ are represented with solid lines, marking the regime in which the young-cluster argument applies.
The $10\,\Myr$ curve is drawn as a dashed grey line because stars with $M_{\mathrm{init}} > 20\,\Msun$ have already died by then and it exits the plotting window below $M_{\mathrm{init}} = 10\,\Msun$. 
The horizontal dotted line marks $F = 0.98$, and the vertical dotted line marks $M_{\mathrm{init}} = 10\,\Msun$.
For all $t \leq 3\,\Myr$, more than $98\%$ of cluster $L_{\mathrm{bol}}$ during the feedback-relevant window comes from stars with $M_{\mathrm{init}} > 10\,\Msun$.}
\label{fig:pms_caveat}
\end{figure*}

\trinity\ starts every simulation with a fully formed stellar population at $t = 0$, placed on the zero-age main sequence (ZAMS): the cluster mass $M_\star = \varepsilon\,\Mcloud$ is distributed across a Kroupa IMF, and \textsc{starburst99} \citep{1999ApJS..123....3L} supplies the feedback tables from that moment onward.
We do not model pre-main-sequence (PMS) evolution.
The error this introduces depends on two things: how long PMS contraction takes for a star of given mass, and which mass range actually generates the feedback.

To answer the first, we use the MESA Isochrones and Stellar Tracks library \citep[MIST v1.2;][]{2016ApJ...823..102C,2016ApJS..222....8D}, built on the \textsc{Mesa} stellar evolution code \citep{2011ApJS..192....3P}.
Models at solar metallicity and zero rotation are adopted to match the \textsc{starburst99} Geneva non-rotating tracks used elsewhere in this paper. The choice is also conservative for the present purpose: changing the rotation prescription affects the detailed massive-star tracks, but the feedback budget still comes from stars that reach the main sequence on timescales short compared to cloud clearing.

Figure~\ref{fig:pms_caveat} (left panel) shows six MIST evolutionary tracks covering $M_{\mathrm{init}} \in [0.5, 60]\,\Msun$, with the PMS segment indicated with dashed lines and ZAMS marked by a filled circle. 
A $60\,\Msun$ star reaches the ZAMS in roughly $17\,\mathrm{kyr}$, a $30\,\Msun$ star in $32\,\mathrm{kyr}$, and a $10\,\Msun$ star in roughly $190\,\mathrm{kyr}$.
A $0.5\,\Msun$ star, by contrast, takes about $160\,\Myr$ to contract, longer than the host GMC itself survives. 
The ZAMS assumption therefore misrepresents low-mass stars; whether that misrepresentation matters depends on how much feedback they produce.

Figure~\ref{fig:pms_caveat} (right panel) plots the cumulative luminosity fraction $F(>M;t)$, i.e. the share of the cluster bolometric luminosity contributed by stars with initial mass above $M$. 
We compute this quantity by integrating MIST isochrones against a \citet{2001MNRAS.322..231K} IMF.
For the stellar mass range used here, the Kroupa IMF is a broken power law,
\begin{equation}
  \xi(M) = k \begin{cases}
    M^{-\alpha^{\mathrm{IMF}}_{1}}, & M_{\mathrm {min}} \leq M < M_{\mathrm {break}}, \\[4pt]
    M_{\mathrm {break}}^{\,\alpha^{\mathrm{IMF}}_{2}-\alpha^{\mathrm{IMF}}_{1}}\, M^{-\alpha^{\mathrm{IMF}}_{2}}, & M_{\mathrm {break}} \leq M \leq M_{\mathrm {max}},
  \end{cases}
  \label{eq:kroupa}
\end{equation}
with $\alpha^{\mathrm{IMF}}_{1} = 1.3$, $\alpha^{\mathrm{IMF}}_{2} = 2.3$, $M_{\mathrm {break}} = 0.5\,\Msun$, $M_{\mathrm {min}} = 0.08\,\Msun$, and $M_{\mathrm {max}} = 120\,\Msun$. Here, 
$k$ is the normalisation factor, and the $M_{\mathrm {break}}^{\,\alpha^{\mathrm{IMF}}_{2}-\alpha^{\mathrm{IMF}}_{1}}$ prefactor enforces continuity at the break. 
The cumulative fraction is then
\begin{equation}
  F(>M; t) \equiv \frac{\displaystyle \int_{M}^{M_\mathrm{to}(t)} L_\mathrm{bol}(M'; t) \, \xi(M') \, \mathrm{d}M'}{\displaystyle \int_{M_\mathrm{min}}^{M_\mathrm{to}(t)} L_\mathrm {bol}(M'; t) \, \xi(M') \, \mathrm{d}M'},
  \label{eq:Fgt}
\end{equation}
where $L_\mathrm{bol}(M; t)$ is the bolometric luminosity of a star of initial mass $M$ at age $t$ read from the MIST isochrone, $M_\mathrm{to}(t)$ is the main-sequence turnoff mass, and the denominator is the total cluster bolometric luminosity.

Note that we use $L_\mathrm{bol}$ as the figure of merit, which is the natural scalar for radiation pressure but not for every feedback channel. 
Photoionisation is driven by the ionising photon rate $\Qi$ and winds by $L_{\mathrm w} = 0.5\,\dot{M}v_\infty^{2}$; each is more concentrated on the massive tail of the IMF than the bolometric luminosity. 
Substituting $Q_{\mathrm{i}}$ or $\Lw$ for $L_{\mathrm{bol}}$ in Eq.~\eqref{eq:Fgt} therefore pushes every entry closer to unity, since stars below $\sim 10\,\Msun$ contribute almost nothing to either channel.

At ages between $0.1$ and $3\,\Myr$ (i.e. the window over which pre-SN feedback clears the natal cloud), more than $98\%$ of the cluster bolometric luminosity comes from stars with $M_{\mathrm{init}} > 10\,\Msun$.
These same stars reach the zero-age main sequence in $\lesssim 0.3\,\Myr$. 
The stars misrepresented by the ZAMS assumption therefore contribute less than $\sim2\%$ of the cluster bolometric luminosity after $\approx0.3\,\Myr$; the ZAMS assumption has a negligible effect on the cluster feedback budget because the low-mass stars it misrepresents are too faint to contribute appreciably.

\section{Constraints on the cloud parameter space}
\label{app:cloud_constraints}

\begin{figure*}[t]
    \centering
    \includegraphics[width=\textwidth]{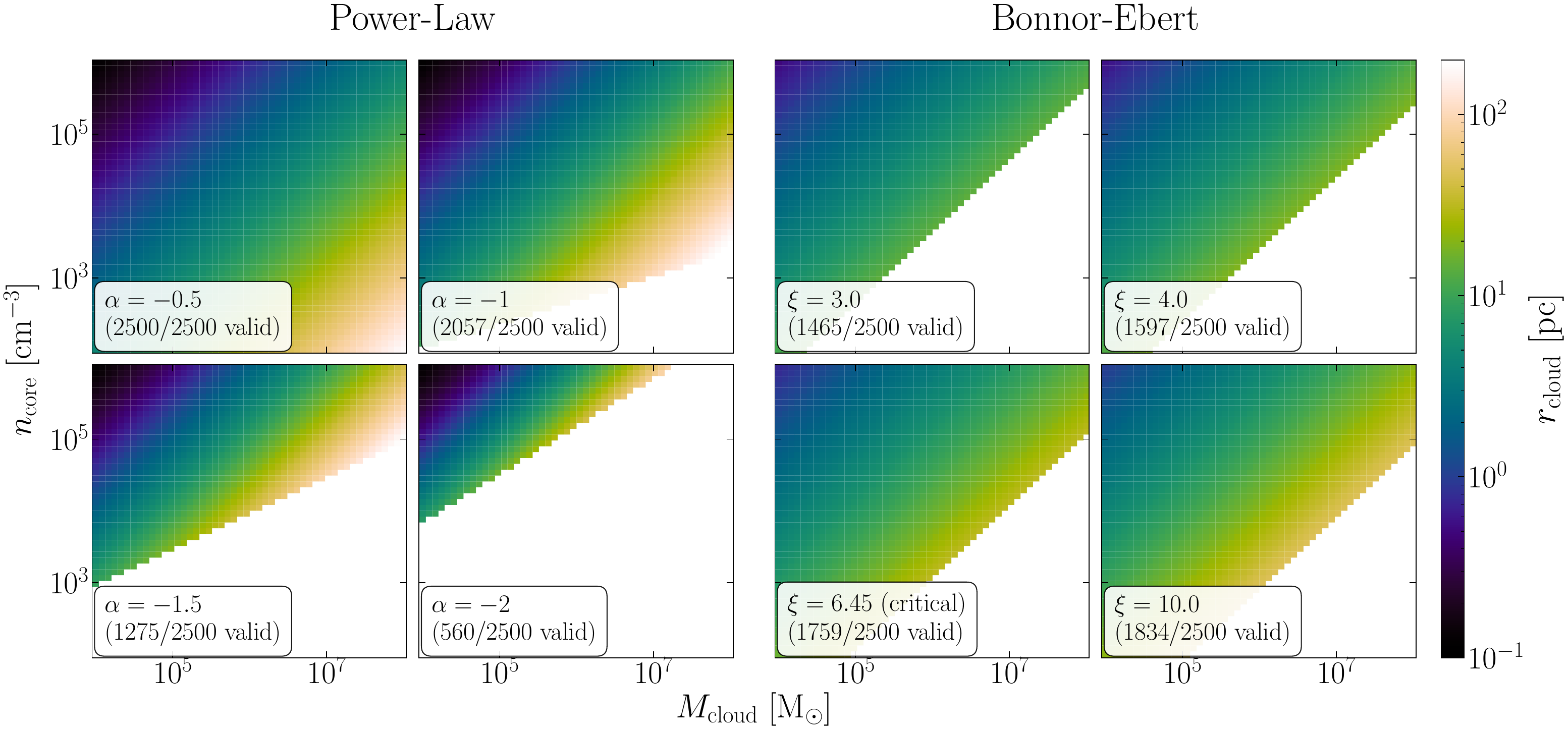}
    \caption{
    \textit{Left:} Permitted parameter space for power-law density profiles ($\rho \propto r^{\alpha_\rho}$) at a fixed core radius $\Rcore = 0.1\,\pc$. 
    Each panel shows the cloud radius $R_\mathrm{cl}$ as a function of cloud mass $\Mcloud$ and core density $\ncore$, for $\alpha_\rho = -0.5$, $-1$, $-1.5$, and $-2$. The white regions are excluded by at least one of two constraints: $R_\mathrm{cl} > 200\,\pc$ or edge density below the ISM value ($n_\mathrm{ISM} = 1\,\ccm$) (see Appendix~\ref{app:cloud_constraints}). 
    The steeper slopes restrict the permitted band to higher core densities at a given cloud mass, because the envelope reaches $n_\mathrm{ISM}$ at a smaller radius. 
    \textit{Right:} Same as in the left panel but for BE spheres with dimensionless radii $\xi_\mathrm{cl} = 3.0$, $4.0$, $6.45$ (critical, marginally stable), and $10.0$ (gravitationally unstable).
    }
    \label{fig:allowed_GMC}
\end{figure*}

Not every combination of cloud parameters ($\Mcloud$, $\ncore$, $\Rcore$, $\alpha_\rho$) yields a physically realisable initial condition. \trinity enforces two constraints before accepting a model. 
First, the cloud radius $R_\mathrm{cl}$ -- fixed by requiring the integrated mass profile to equal the prescribed cloud mass $\Mcloud$ -- must remain within the observed range of Galactic GMC and molecular-cloud-complex sizes. 
We adopt $R_\mathrm{cl} \le 200\,\pc$ as a conservative upper cutoff, motivated by the scale of the largest Galactic cloud complexes \citep{1987ApJ...319..730S,2017ApJ...834...57M}. Second, the density at the cloud boundary must remain above the ambient ISM value, $n(R_\mathrm{cl}) \ge n_\mathrm{ISM}$. For power-law profiles with slope $\alpha_\rho < 0$, this condition gives
\begin{equation}
    R_\mathrm{cl}/\Rcore \le (\ncore/n_\mathrm{ISM})^{1/|\alpha_\rho|} \,,
    \label{eq:edge_density_bound}
\end{equation}
which becomes increasingly restrictive for steeper slopes.

Figure~\ref{fig:allowed_GMC} maps the permitted parameter space in the ($\Mcloud$, $\ncore$) plane for power-law and BE profiles. 
For a fixed core radius $\Rcore = 0.1\,\pc$, valid combinations form a diagonal band whose width and position depend on the profile shape. Steeper power-law slopes and larger BE dimensionless radii $\xi_\mathrm{cl}$ restrict the permitted region to higher core densities at a given cloud mass, because the extended envelope reaches $n_\mathrm{ISM}$ at a smaller physical radius. 
Observational surveys of Galactic molecular clouds find mean densities $\bar{n} \sim 10^{2}$--$10^{3}\,\ccm$ on GMC scales \citep{1987ApJ...319..730S,2009ApJ...699.1092H,2017ApJ...834...57M}, rising to $\gtrsim 10^{4}\,\ccm$ in dense clumps and cores \citep{2007ARA&A..45..339B}. 
Massive star-forming clumps and cores often show approximate power-law density structures, with typical indices near $p\sim 1.5$--$2$ for $n\propto r^{-p}$, albeit with substantial source-to-source scatter and modelling systematics \citep{2002ApJ...566..945B,2002ApJS..143..469M}.

\section{The cloud--ISM density discontinuity}
\label{app:density_smoothing}

In many idealised cloud models, the cloud--ISM transition is treated as a step function at $r=R_\mathrm{cloud}$ \citep[e.g.][]{2017MNRAS.470.4453R}. 
In our model grid this jump often spans several orders of magnitude in density; thus, the solver may attempt to reduce the step size excessively and terminate.
Whether this happens depends on the shell speed at the crossing. 
In our tests, the failure appears when the shell reaches the cloud edge at $\gtrsim 10\,\kms$: a slow shell traverses the density jump in many small accepted steps, whereas a fast shell encounters the jump too abruptly for the adaptive integrator to regularise. 

We avoid this numerical error by replacing the discontinuity with a hyperbolic-tangent bridge across $R_\mathrm{cloud}$,
\begin{equation}
\label{eq:tanh_bridge}
w(r)=\frac{1}{2}\left[1+\tanh\left(\frac{r-R_\mathrm{cloud}}{\delta_\mathrm{smooth}}\right)\right],
\end{equation}
with width $\delta_\mathrm{smooth}=0.01\,R_\mathrm{cloud}$, and defining the smoothed profile as
\begin{equation}
\label{eq:tanh_bridge_2}
n_\mathrm{smooth}(r)=n_\mathrm{cloud}(r)\,[1-w(r)] + n_\mathrm{ISM}\,w(r),
\end{equation}
where $n_\mathrm{cloud}$ is the density profile within the cloud. Observed cloud boundaries are structured \citep{1991ApJ...378..186F}, so we use this bridge as a numerical regularisation only.

\end{appendix}

\end{document}